\pdfoutput=1
\documentclass[pre,twocolumn,showpacs,showkeys,superscriptaddress,amsmath,amssymb,floatfix]{revtex4-1}

\usepackage{graphicx}
\usepackage{bm}
\usepackage{placeins}
\usepackage{graphics}
\usepackage{color}
\usepackage{multirow}
\usepackage{blindtext}
\usepackage[colorlinks=true, linkcolor=blue, urlcolor=blue,
citecolor=blue]{hyperref}

\begin{document}

\title{Scattering-induced and highly tunable by gate damping-like spin-orbit torque in  graphene doubly proximitized by two-dimensional magnet Cr$_2$Ge$_2$Te$_6$ and  WS$_2$}

%

\author{Klaus Zollner}
\email{klaus.zollner@physik.uni-regensburg.de}
\affiliation{Institute for Theoretical Physics, University of Regensburg, 93040 Regensburg, Germany}
             
\author{Marko D. Petrovi\'{c}}
\affiliation{Department of Mathematical Sciences, University of Delaware, Newark, DE 19716, USA}
\affiliation{Department of Physics and Astronomy, University of Delaware, Newark, DE 19716, USA}

\author{Kapildeb Dolui}
\affiliation{Department of Physics and Astronomy, University of Delaware, Newark, DE 19716, USA}
                          
\author{Petr Plech\'a\v{c}}
\affiliation{Department of Mathematical Sciences, University of Delaware, Newark, DE 19716, USA}             
             
\author{Branislav K. Nikoli\'{c}}
\email{bnikolic@udel.edu}
\affiliation{Department of Physics and Astronomy, University of Delaware, Newark, DE 19716, USA}

\author{Jaroslav Fabian}
\affiliation{Institute for Theoretical Physics, University of Regensburg, 93040 Regensburg, Germany}

\begin{abstract}
Graphene sandwiched between semiconducting monolayers of ferromagnet Cr$_2$Ge$_2$Te$_6$ and transition-metal  dichalcogenide  WS$_2$ acquires both spin-orbit (SO), of valley-Zeeman and Rashba types, and exchange couplings. Using first-principles  combined with quantum transport calculations, we predict that such doubly proximitized graphene within van der Waals heterostructure will exhibit SO torque driven by unpolarized charge current. This system lacking spin Hall current, putatively considered to be necessary for efficient damping-like (DL) SO torque that plays a key role in magnetization switching, demonstrates how DL torque component  can be generated {\em solely} by skew-scattering off {\em  spin-independent} potential barrier or impurities in {\em purely}  two-dimensional electronic transport due to the presence of proximity SO coupling and its spin texture tilted out-of-plane. This leads to current-driven nonequilibrium spin density emerging in {\em all spatial directions}, whose cross product with proximity magnetization yields DL SO torque, unlike the ballistic regime with no scatterers in which {\em only} field-like (FL) SO torque appears.  In contrast to SO torque on conventional metallic ferromagnets in contact with three dimensional SO-coupled  materials, the ratio of FL and DL torque can be tuned by more than an order of magnitude via combined top and back  gates.
\end{abstract}

\maketitle


\section{Introduction}

The spin-orbit (SO) torque~\cite{Manchon2019} is a phenomenon in which unpolarized charge current injected 
parallel to the interface of a bilayer of ferromagnetic metal (FM) and SO-coupled material induces magnetization dynamics of the FM layer. 
For many possible applications of torque~\cite{Locatelli2014}, such as nonvolatile magnetic random access
memories~\cite{Ramaswamy2018} or artificial neural networks~\cite{Borders2017}, it is crucial to switch magnetization 
from up to down along the direction perpendicular to the interface. This has led to intense search for optimal 
SO-coupled materials and their interfaces with FM layers which yield large SO torque while using as small as possible 
injected current. Thus far, minimal current density $j$ for magnetization switching has been achieved using topological insulators 
($j \sim 6\,\times\,10^5$\,A/cm$^2$~\cite{Wang2017}) and Weyl semimetals ($j \sim 3\,\times\,10^5$\,A/cm$^2$~\cite{Shi2019}), which is 
two orders of magnitude smaller than $j$ required in early SO torque-operated devices employing heavy metals~\cite{Manchon2019,Zhu2019a}.

\begin{figure}
	\includegraphics[scale=0.95]{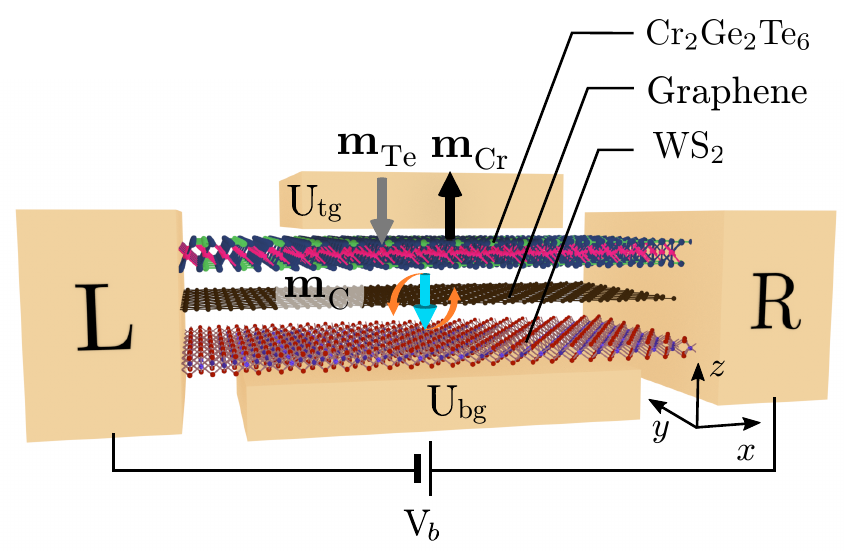}
	\caption{Schematic view of Cr$_2$Ge$_2$Te$_6$/graphene/WS$_2$ vdW heterostructure attached to macroscopic left and right reservoirs with a small
		bias voltage ${\rm V}_b$ between them injecting unpolarized charge current into graphene layer. A back gate voltage $U_{\rm bg}$ and a top gate voltage $U_{\rm tg}$ applied over a smaller ``active region'' are assumed to control the Fermi energy and the local on-site potential in graphene, respectively. The unit vectors of magnetic moments on Cr, Te and C atoms are denoted as $\mathbf{m}_\mathrm{Cr}$, $\mathbf{m}_\mathrm{Te}$ and $\mathbf{m}_\mathrm{C}$, respectively, where only $\mathbf{m}_\mathrm{C}$ experiences SO torque-driven dynamics.}  \label{fig:fig1}
\end{figure}

Further optimization could be achieved by using van der Waals (vdW) heterostructures~\cite{Dolui2020} of very recently discovered two-dimensional (2D) ferromagnets~\cite{Gibertini2019,Cortie2019} and 2D SO-coupled materials where current flows only through few monolayers and no 
Joule heat is wasted by its flow through the bulk. Furthermore, vdW heterostructures offer atomically flat and highly transparent interfaces, as well as possibility to use external manipulations---such as gating, straining and controlling coupling between 2D materials---in order to tune~\cite{Lv2018} the magnitude and ratio of field-like (FL) and damping-like (DL) components of SO torque. The traditional labeling of torque components stems from how they affect dynamics of  magnetization $\mathbf{m}_\mathrm{C}$ viewed as a classical vector---FL torque changes precession around an effective magnetic field (such as  along the $y$-axis in the case of vdW heterostructure in Fig.~\ref{fig:fig1}), while DL torque competes with damping and  plays a key role~\cite{Yan2015a} in magnetization switching (such by directing magnetization $\mathbf{m}_\mathrm{C}$ towards the $y$-axis in Fig.~\ref{fig:fig1}). Controlling their ratio and relative sign is of great interest for applications since it makes it possible to tailor the switching probability~\cite{Yoon2017}.

The {\em critical problem} in the field of SO torques---understanding of competing microscopic mechanisms behind these components, especially the DL one,  and control of their magnitude and ratio---remains unresolved. For example, experiments on FM/heavy-metal bilayers suggest~\cite{Zhu2019a} that DL SO torque is primarily generated by the spin Hall current~\cite{Sinova2015} from the bulk of a heavy metal, while interfacial SO coupling (SOC) is  detrimental because it generates spin memory loss~\cite{Dolui2017,Belashchenko2016,Gupta2020} and, therefore, reduction of spin Hall current. Conversely, first-principles quantum transport calculations~\cite{Belashchenko2020,Mahfouzi2020} on FM/heavy-metal bilayers find that interfacial SOC and spin Hall current can equally contribute to total DL SO torque. The purely interfacial contribution is easily identified in experiments~\cite{Luo2019} and computational studies~\cite{Belashchenko2020,Mahfouzi2020} as the one that is {\em independent} on the thickness of heavy-metal layer, but its microscopic mechanisms remain poorly understood. Although reflection and transmission of electron spins {\em traversing} SO-coupled interface can lead to interface-generated spin currents inducing DL SO torque~\cite{Kim2017,Amin2018}, both this mechanism and the spin Hall current are {\em inoperative} in 2D transport within vdW heterostructures or in FM/topological-insulator bilayers~\cite{Wang2017,Ghosh2018}. Thus far, it has been argued that {\em spin-independent} impurities {\em cannot} generate DL torque in purely 2D transport~\cite{Pesin2012a,Qaiumzadeh2015,Ado2017}, as well as that the same impurities can completely cancel~\cite{Qaiumzadeh2015,Ado2017} the intrinsic Berry curvature mechanism~\cite{Kurebayashi2014} of DL SO torque as the only other possibility discussed for purely 2D transport.

\begin{figure}
	\includegraphics[width=0.99\columnwidth]{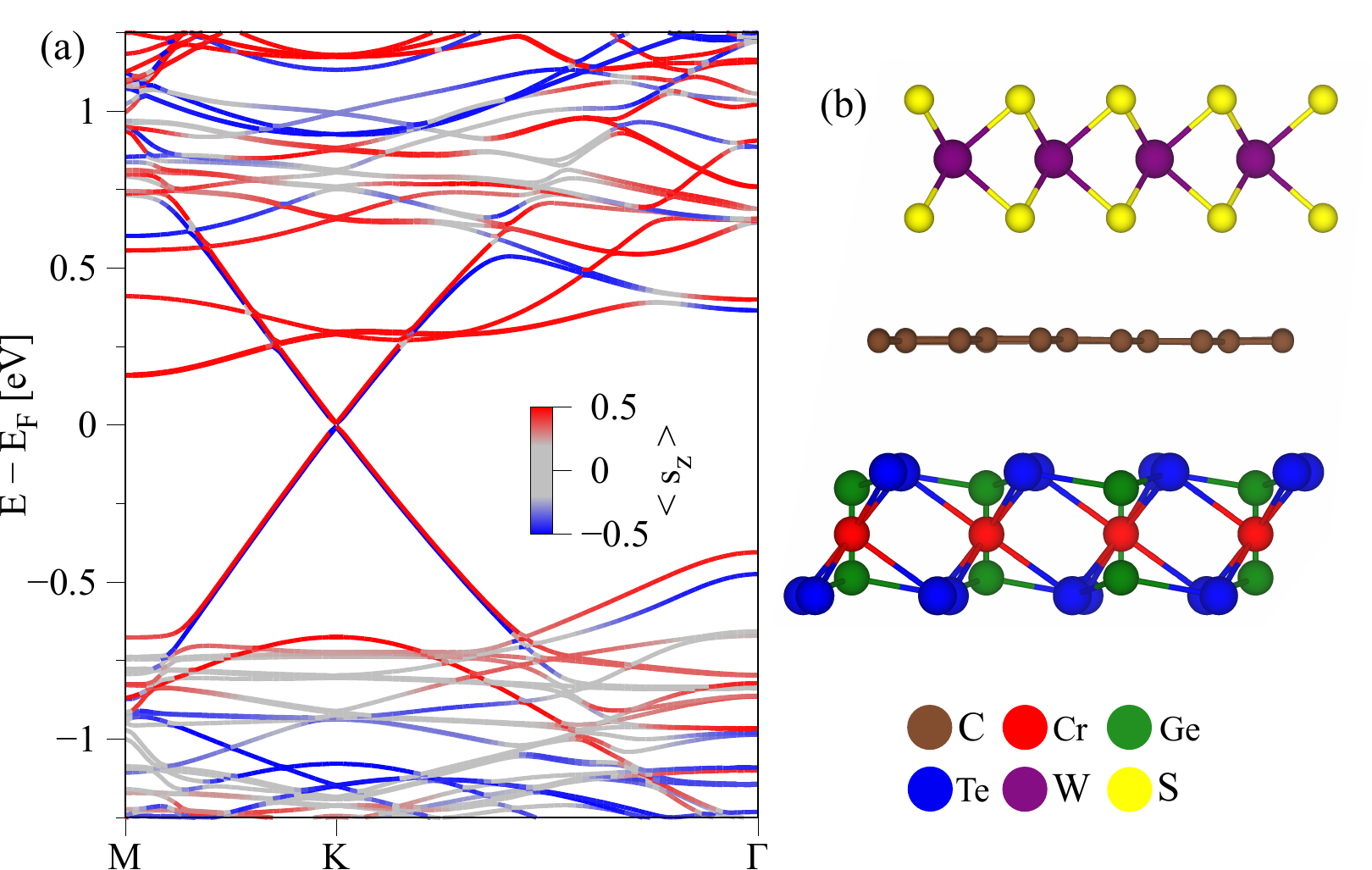}
	\caption{(a) The DFT-calculated electronic band structure of graphene
		sandwiched between monolayer Cr$_2$Ge$_2$Te$_6$ and monolayer
		WS$_2$. The color corresponds to the $\langle \hat{s}_z \rangle$
		expectation value in equilibrium. (b) Side view of the geometry
		of the trilayer supercell.} 
	\label{fig:fig2} 
\end{figure}

In this study, we design a realistic (i.e., built atom-by-atom) vdW heterostructure monolayer-Cr$_2$Ge$_2$Te$_6$/graphene/monolayer-WS$_2$, as depicted in Fig.~\ref{fig:fig1}. Here  Cr$_2$Ge$_2$Te$_6$~\cite{Menichetti2019,Carteaux1995} is an example of the recently discovered 2D ferromagnets~\cite{Gibertini2019,Cortie2019}, and  WS$_2$ is a transition-metal dichalcogenide (TMD) with strong SOC in monolayer form~\cite{Gmitra2016}. Unlike isolated graphene which is nonmagnetic and it hosts minuscule intrinsic SOC~\cite{Gmitra2009}, {\em doubly proximitized} graphene within Cr$_2$Ge$_2$Te$_6$/graphene/WS$_2$ trilayer offers a versatile ``theoretical laboratory'' to differentiate between competing mechanisms of  SO torque and thereby learn how to control them. This can be accomplished by switching on and off different terms in the first-principles calculations [Figs.~\ref{fig:fig2} and ~\ref{fig:fig3}] derived Hamiltonian [Eq.~\eqref{eq:hamiltonian}] of the heterostructure and then by performing quantum transport calculations.

The continuous version of such a Hamiltonian [Eq.~\eqref{eq:hamiltonian}] can be used as an input for diagrammatic perturbative  analytic calculations~\cite{Qaiumzadeh2015,Ado2017,Sousa2020} based on the Kubo formula~\cite{Freimuth2014} for nonequilibrium spin density $\langle \hat{\mathbf{s}} (\mathbf{r})\rangle_\mathrm{CD}$  within graphene, which exerts torque  $\propto \langle \hat{\mathbf{s}} (\mathbf{r}) \rangle_\mathrm{CD} \times \mathbf{m}_\mathrm{C}(\mathbf{r})$~\cite{Dolui2020,Nikolic2018,Dolui2020a,Belashchenko2019,Belashchenko2020} on the local magnetization $\mathbf{m}_\mathrm{C}(\mathbf{r})$ in graphene. Note that  $\mathbf{m}_\mathrm{C}(\mathbf{r})$ is induced by magnetic proximity effect originating from Te atoms of Cr$_2$Ge$_2$Te$_6$, so that  $\mathbf{m}_\mathrm{C} \parallel \mathbf{m}_\mathrm{Te}$ but these two magnetic moments are antiparallel to $\mathbf{m}_\mathrm{Cr}$ on Cr atoms, as  illustrated in Fig.~\ref{fig:fig1}. 

Alternatively, one can discretize continuous Hamiltonian in Eq.~\eqref{eq:hamiltonian} to obtain its TB version, as provided in Appendix~\ref{sec:tbh}, and combine it with computational quantum transport. We feed such first-principles TB Hamiltonian into the calculations~\cite{Nikolic2018,Dolui2020a,Belashchenko2019} of current-driven (CD) part of nonequilibrium density matrix $\hat{\rho}_\mathrm{CD}$, which yields \mbox{$\langle \hat{\mathbf{s}}_i \rangle_\mathrm{CD}=\mathrm{Tr}_\mathrm{spin}\, [\hat{\rho}_\mathrm{CD} \hat{\mathbf{s}}]$} on site $i$ of TB lattice for \mbox{$\hat{\mathbf{s}}=(\hat{s}_x,\hat{s}_y,\hat{s}_z)$} as the vector of the Pauli matrices and the trace being performed only in the spin space. The matrix $\hat{\rho}_\mathrm{CD}$ is computed via the nonequilibrium Green function (NEGF) formalism~\cite{Stefanucci2013} for the Landauer setup where vdW heterostructure in Fig.~\ref{fig:fig1} is divided into the left (L) semi-infinite-lead  connected to ``active region'' which is connected to the right (R) semi-infinite-lead. Both semi-infinite leads and the active region in Fig.~\ref{fig:fig1} are made of the same trilayer, assumed to be also infinite in the transverse $y$-direction. The leads terminate into the L and R macroscopic reservoirs kept at electrochemical potential difference $\mu_\mathrm{L}-\mu_\mathrm{R} = eV_\mathrm{b}$ with small bias voltage $V_\mathrm{b}$ in the linear-response regime. Such nonperturbative (i.e., numerically exact) approach can be used to investigate SO torque in the {\em diffusive} transport regime~\cite{Belashchenko2019,Belashchenko2020}; as well as in the {\em ballistic}~\cite{Chang2015,Kalitsov2017} and {\em quasiballistic} regimes which are not accessible in the Kubo-formula-based calculations requiring nonzero electric field and momentum relaxation mechanisms. By using homogeneous potential barrier in the active region due to the top electrostatic gate~\cite{Young2009,Liu2012d} in Fig.~\ref{fig:fig1}, and/or short-ranged impurities generating spin-independent random potential~\cite{Sousa2020,Milletari2017}, we demonstrate [Fig.~\ref{fig:fig4}] our {\em principal result---spin-independent scatterers can be a sole generator of nonzero DL SO torque in purely 2D transport}, so they can be  manipulated~\cite{Sousa2020} to tune its magnitude.

The paper is organized as follows. Section~\ref{sec:dft} explains our density functional theory (DFT) calculations for Cr$_2$Ge$_2$Te$_6$/graphene/WS$_2$ trilayer, with additional details and cases (such as Cr$_2$X$_2$Te$_6$/graphene/WS$_2$ trilayers and Cr$_2$X$_2$Te$_6$/graphene bilayers where X = \{Si,Ge,Sn\})  provided in Appendices~\ref{sec:detailsdft}--\ref{sec:cgtgraphene}. Section~\ref{sec:dft} also provides a continuous effective Hamiltonian of graphene with parameters derived from DFT calculations, with its tight-binding (TB) version given in Appendix~\ref{sec:tbh}. The quantum transport calculations of current-driven  nonequilibrium spin densities and SO torques, based on the TB Hamiltonian (Appendix~\ref{sec:tbh}) of doubly proximitized graphene with additional gate- or disorder-induced on-site potential, are presented in Sec.~\ref{sec:sot}. We conclude in Sec.~\ref{sec:conclusions}.

\section{First-principles band structure and thereby derived low-energy continuous Hamiltonian}\label{sec:dft}

\begin{figure}[t]
	\includegraphics[width=0.99\columnwidth]{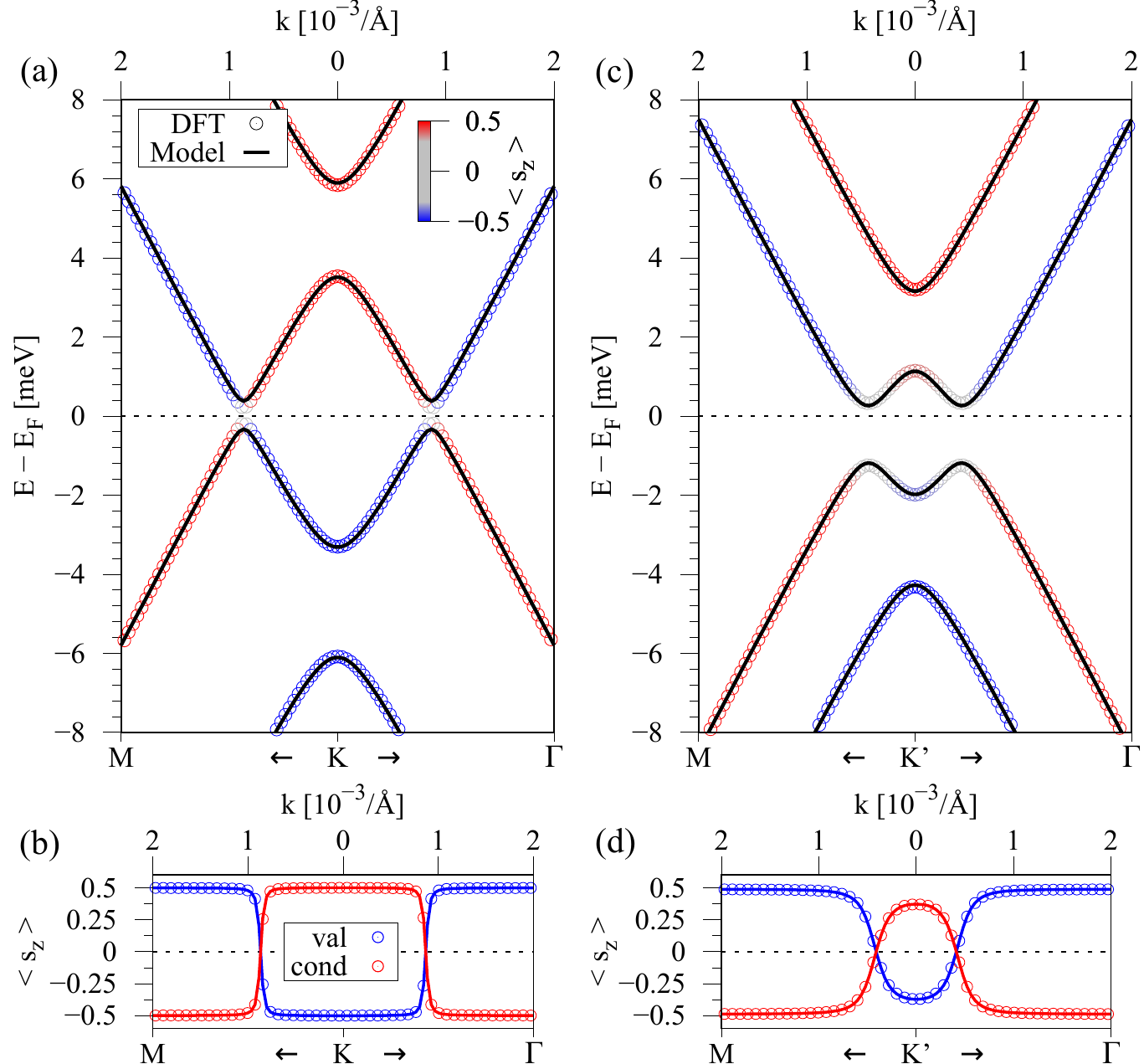}
	\caption{(a) Zoom to the low energy bands around $E-E_F=0$ in
		Fig.~\ref{fig:fig2}. The Dirac cone is split by proximity
		induced exchange interaction and SOC. Color of the bands
		corresponds to $\langle \hat{s}_z \rangle$ expectation value in
		equilibrium. (b) The $\langle \hat{s}_z \rangle$ expectation value of
		the valence (conduction) band in blue (red). Symbols are
		first-principles computed data and solid lines are fits using
		the continuous Hamiltonian in Eq.~\eqref{eq:hamiltonian}.
		(c), (d) The same information as in panels (a), (b), but for
		K$^\prime$ point.} \label{fig:fig3}
\end{figure}

\begin{figure}
	\includegraphics[scale=1.0]{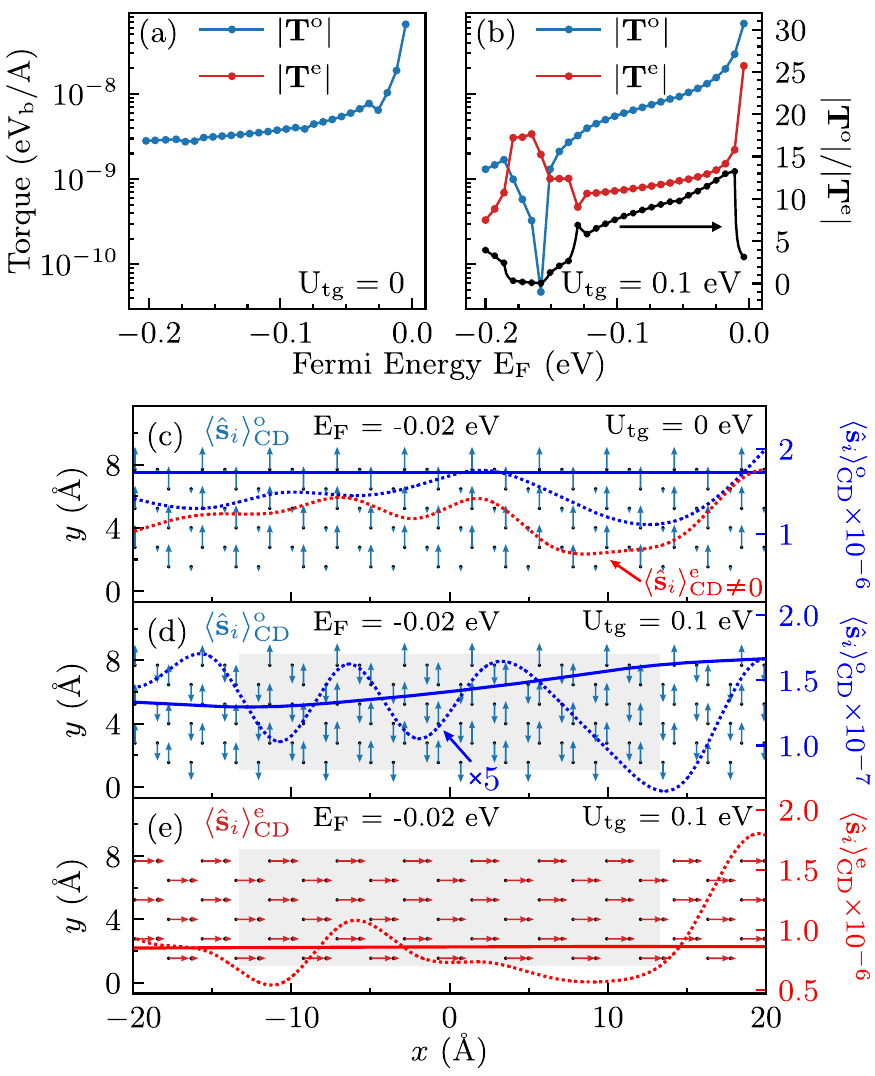}
	\caption{The magnitude of odd $|\mathbf{T}^{\rm o}|$ and even $|\mathbf{T}^{\rm e}|$  SO torque components as a function of $E_F$ for proximity magnetization of graphene $\mathbf{m}_\mathrm{C} \parallel \hat{\mathbf{z}}$ [Fig.~\ref{fig:fig1}]: (a) without the  on-site energy $U_{\rm tg} = 0$; and (b) with a constant on-site energy $U_{\rm tg} = 0.1$~eV in the active region which creates a homogeneous potential barrier. The corresponding odd and even components of nonequilibrium spin density $\langle \hat{\mathbf{s}}_{i}\rangle_{\rm CD}^\mathrm{e,o}$ [Eq.~\eqref{eq:sot}] at \mbox{${E}_F = -0.02$~eV} are shown as vector fields in: (c)  without the barrier; and (d),(e) with barrier $U_{\rm tg} = 0.1$~eV within the gray rectangle. Solid lines in (c)--(d) show the magnitude of the sum of all arrows at coordinate $x$, while the dashed lines show how impurities modify the solid lines. The spin-independent impurities are modeled by  random on-site energies $\in [-V/2,V/2]$ drawn from a uniform distribution with \mbox{$V=3$ eV} to ensure the diffusive transport regime (characterized by the shot noise Fano factor $F \simeq 1/3$~\cite{Lewenkopf2008}). Each dashed line is geometric average over 20 disorder realizations.}
	\label{fig:fig4}
\end{figure}

Using DFT, we first compute  the band structure [Figs.~\ref{fig:fig2} and ~\ref{fig:fig3}] of Cr$_2$Ge$_2$Te$_6$/graphene/WS$_2$ trilayer. Since the Dirac cones of graphene are largely preserved within the band gap of  Cr$_2$Ge$_2$Te$_6$, an effective  Hamiltonian for graphene {\em only} can be extracted in  Eq.~\eqref{eq:hamiltonian}. The DFT calculations are performed on the supercell  [Fig.~\ref{fig:fig2}(b)] of Cr$_2$Ge$_2$Te$_6$/graphene/WS$_2$ trilayer using  {\tt Quantum ESPRESSO}~\cite{Giannozzi2009} and {\tt WIEN2k} packages~\cite{Blaha2001}, with additional details provided in Sec.~\ref{sec:detailsdft}. The band structure in  Fig.~\ref{fig:fig2}(a) shows that the Dirac cone of graphene is preserved and located in the global band gap, so it can be probed by charge and spin
transport. Figure~\ref{fig:fig3} shows a zoom to the fine
structure around K and K$^\prime$ points with a fit to our
continuous Hamiltonian
\begin{subequations}
	\label{eq:hamiltonian}
	\begin{eqnarray}
	\hat{H} & = & 
	\hat{H}_{0} + 
	\hat{H}_{\Delta} +
	\hat{H}_{\textrm{I}} + 
	\hat{H}_{\textrm{R}} +
	\hat{H}_{\textrm{ex}} + 
	\hat{H}_{\xi}, \\
	\hat{H}_{0} & = & 
	\hbar v_{\textrm{F}} 
	(\tau k_x \hat{\sigma}_x - k_y \hat{\sigma}_y)
	\otimes \hat{s}_0, \\
	\hat{H}_{\Delta} & = & 
	\Delta \hat{\sigma}_z \otimes \hat{s}_0,\\
	\hat{H}_{\textrm{I}} & = & 
	\tau (\lambda_{\textrm{I}}^\textrm{A} \hat{\sigma}_{+}
	+\lambda_{\textrm{I}}^\textrm{B} \hat{\sigma}_{-})
	\otimes \hat{s}_z,\\
	\hat{H}_{\textrm{R}} & = &
	-\lambda_{\textrm{R}}
	(\tau \hat{\sigma}_x \otimes \hat{s}_y +
	\hat{\sigma}_y \otimes \hat{s}_x),\\
	\hat{H}_{\textrm{ex}} & = & 
	(-\lambda_{\textrm{ex}}^\textrm{A} \hat{\sigma}_{+}
	+\lambda_{\textrm{ex}}^\textrm{B} \hat{\sigma}_{-}) 
	\otimes \hat{s}_z,\\
	\hat{H}_{\xi} & = & 
	\tau \xi \hat{\sigma}_0 \otimes \hat{s}_0.
	\end{eqnarray}
\end{subequations}
This is derived from first-principles calculations, as well as additional symmetry arguments~\cite{Kochan2017,Phong2017}, and it 
is valid in the vicinity of both Dirac points, K and K$^\prime$. It reproduces the gap at the Dirac point 
and the spin-splitting of the bands due to proximity SOC~\cite{Marmolejo-Tejada2017,Zutic2019}, of valley-Zeeman and Rashba types~\cite{Gmitra2015,Gmitra2016,Kochan2017},  
which originates from WS$_2$; as well as due to proximity magnetism~\cite{Hallal2017} from Cr$_2$Ge$_2$Te$_6$ into graphene. These interactions are captured by relevant terms in our continuous Hamiltonian [Eq.~\eqref{eq:hamiltonian}], with parameters provided in Table~\ref{tab:table1}. 

The vdW heterostructure we consider has broken time-reversal symmetry and only $C_3$ symmetry. We denote $v_{\textrm{F}}$ as
the Fermi velocity, and wavevector components $k_x$ and $k_y$ are
measured from $\pm$K. The valley index is $\tau = \pm 1$ for $\pm$K and pseudospin matrices are $\hat{\sigma}_i$,
acting on sublattice space (C$_\textrm{A}$, C$_\textrm{B}$), with
$i = \{ 0,x,y,z \}$ where $i = 0$ denotes a unit $2 \times 2$
matrix. For notational convenience, we use $\hat{\sigma}_{\pm} =
\frac{1}{2}(\hat{\sigma}_z \pm \hat{\sigma}_0)$. The staggered
potential gap is $\Delta$, and the parameters
$\lambda_{\textrm{I}}^\textrm{A}$ and
$\lambda_{\textrm{I}}^\textrm{B}$ denote the sublattice-resolved
intrinsic SOC. The parameter $\lambda_{\textrm{R}}$ is the Rashba 
SOC, and the proximity exchange interaction parameters are
$\lambda_{\textrm{ex}}^\textrm{A}$ and
$\lambda_{\textrm{ex}}^\textrm{B}$. The parameter $\xi$ describes
valley exchange coupling resulting from an in-plane magnetization
component~\cite{Phong2017}. The four basis states are
$|\Psi_{\textrm{A}}, \uparrow\rangle$,  $|\Psi_{\textrm{A}},
\downarrow\rangle$,  $|\Psi_{\textrm{B}}, \uparrow\rangle$,  and
$|\Psi_{\textrm{B}}, \downarrow\rangle$.  The continuous  
Hamiltonian is centered around the Fermi level at zero energy.
Since first-principles results capture doping effects, we also introduce 
parameter $E_\mathrm{D}$ (termed Dirac point energy) which shifts the global band structure.

The fitting parameters for Cr$_2$Ge$_2$Te$_6$/graphene/WS$_2$ trilayer 
are summarized in Table~\ref{tab:table1}. They are in agreement
with previous calculations for individual graphene/TMD~\cite{Gmitra2015,Gmitra2016} and
graphene/Cr$_2$Ge$_2$Te$_6$ slabs~\cite{Zhang2015a}.
As seen from Fig.~\ref{fig:fig3},  
Hamiltonian in Eq.~\eqref{eq:hamiltonian} with parameters from
Tab.~\ref{tab:table1} perfectly reproduces first-principles results, including the spin $\langle \hat{s}_z \rangle$ expectation values in
both valleys. The valley degeneracy is broken, which is
best manifested at the highest (lowest) spin-$\uparrow$ (spin-$\downarrow$) bands at K and K$^\prime$. This is due to the interplay of
the proximity exchange and SO couplings which splits the bands in the two valleys differently. Furthermore, the Rashba SOC mixes the
spin states and opens a global band gap, different for the two
valleys. Appendix~\ref{sec:spintexture} also provides equilibrium spin textures within Cr$_2$Ge$_2$Te$_6$/graphene/WS$_2$ trilayer. For comparison, band structure and fitting parameters for other Cr$_2$X$_2$Te$_6$/graphene/WS$_2$ trilayers and Cr$_2$X$_2$Te$_6$/graphene bilayers with X = \{Si,Ge,Sn\} are 
provided in Appendix~\ref{sec:cgtgraphene}.

\begin{figure}
	\includegraphics[width=0.99\columnwidth]{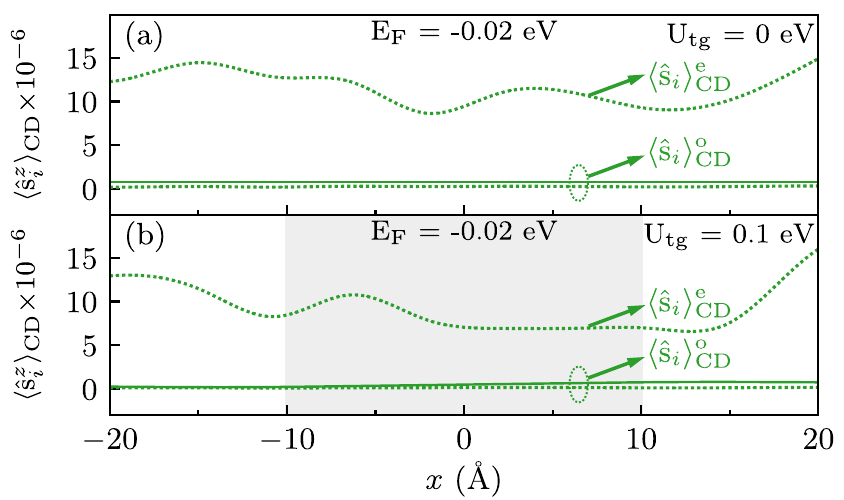}
	\caption{Out-of-plane (i.e.,  along the $z$-axis in the coordinate system of Fig.~\ref{fig:fig1})  component of CD nonequilibrium spin density $\langle \hat{s}^{z}_i\rangle_{\rm CD}$ in doubly proximitized graphene with: (a) no potential barrier $U_{\rm tg} = 0$~eV; or (b) with potential barrier due to  
		homogeneous on-site energy $U_{\rm tg} = 0.1$~eV within the region marked by gray rectangle. Solid lines are for clean graphene, while dashed line include spin-independent impurities modeled by random on-site energies $\in [-V/2,V/2]$ drawn from a uniform distribution with \mbox{$V=3$ eV} to ensure diffusive transport regime. Each dashed line is geometric average over 20 disorder realizations. The magnetization $\mathbf{m}_{\rm C}$ is oriented along the $z$-axis and the Fermi energy is \mbox{$E_F = -0.02$ eV}. The corresponding in-plane components of  $\langle \hat{\mathbf{s}}_i\rangle_{\rm CD}$ are shown in Fig.~\ref{fig:fig4}(c)--(e). The arrows denote how different $\langle \hat{s}^{z}_i\rangle_{\rm CD}$ curves contribute to $\langle \hat{\mathbf{s}}_i\rangle_{\rm CD}^\mathrm{e}$ or $\langle \hat{\mathbf{s}}_i\rangle_{\rm CD}^\mathrm{o}$ vectors.}
	\label{fig:sz}
\end{figure}
%


\section{Quantum transport calculations of SO torque}\label{sec:sot}
	
The NEGF is computed for the active region of vdW heterostructure in Fig.~\ref{fig:fig1}. Since current flows only through graphene, the active region is 
modeled as graphene nanoribbon with armchair edges that  is composed of $N_a=6$ dimer lines across the ribbon width $W$. The nanoribbon is described by the first-principles TB Hamiltonian [Eq.~\eqref{eq:fptbh}].  To remove the effect of the edges and mimic wide junctions along the $y$-axis often employed experimentally~\cite{Young2009}, sites on the lower and upper edge of the nanoribbon are connected~\cite{Liu2012d} by a hopping with $e^{ik_yW}$ phase, where $k_y \in [-\pi/W, \pi/W)$. Thus, all transport quantities are then  $k_y$-points sampled to take into account an infinitely wide system~\cite{Liu2012d}. 

The conventional spin-transfer torque~\cite{Nikolic2018} in spin valves and magnetic tunnel junctions is decomposed into the FL and DL components, $\mathbf{T}=\mathbf{T}^\mathrm{DL} + \mathbf{T}^\mathrm{FL}$, each of which has relatively simple angular dependence. Conversely, the SO torque with its complex angular dependence~\cite{Garello2013} is naturally decomposed~\cite{Belashchenko2019,Belashchenko2020}, $\mathbf{T}=\mathbf{T}^\mathrm{e} + \mathbf{T}^\mathrm{o}$, into the odd $\mathbf{T}^\mathrm{o}$ and  even $\mathbf{T}^\mathrm{e}$ components in the magnetization $\mathbf{m}_\mathrm{C}$ receiving the torque. They are obtained by averaging over $N_{\rm A}$ atoms of the triangular sublattice  A and $N_{\rm B}$ atoms of the triangular sublattice B of graphene
\begin{eqnarray}\label{eq:sot}
\mathbf{T}^\mathrm{e,o}(k_y) & = &
\frac{1}{N_{\rm A}}\sum_{i \in {\rm A}}
\left(-\frac{2\lambda^{\rm A}_{\rm ex}}{\hbar} \right)
\langle \hat{\mathbf{s}}_{i}\rangle_{\rm CD}^\mathrm{e,o}(k_y)
\times \mathbf{m}_\mathrm{C} \nonumber \\
&& +  \frac{1}{N_{\rm B}}\sum_{j \in {\rm B}}
\frac{2\lambda^{\rm B}_{\rm ex}}{\hbar}\langle 
\hat{\mathbf{s}}_{j}\rangle_{\rm CD}^\mathrm{e,o}(k_y)
\times \mathbf{m}_\mathrm{C}.
\end{eqnarray}
The total torque is the sum over the first Brillouin zone (BZ), \mbox{$\mathbf{T}^\mathrm{e,o} = \frac{W}{2\pi} \int_{\rm BZ} \mathbf{T}^\mathrm{e,o}(k_y) dk_y$}. The NEGF-based algorithm to split the density matrix, $\hat{\rho}_\mathrm{CD} = \hat{\rho}_\mathrm{CD}^\mathrm{e} + \hat{\rho}_\mathrm{CD}^\mathrm{o}$, is given in Refs.~\cite{Nikolic2018,Dolui2020a}, so that tracing Pauli matrices with $\hat{\rho}_\mathrm{CD}^\mathrm{e}$ and $\hat{\rho}_\mathrm{CD}^\mathrm{o}$ yields directly $\langle \hat{\mathbf{s}}_{i}\rangle_{\rm CD}^\mathrm{e}$ and $\langle \hat{\mathbf{s}}_{i}\rangle_{\rm CD}^\mathrm{o}$ and the corresponding SO torque components in Eq.~\eqref{eq:sot}.

The magnitudes of odd and even SO torque components are plotted in Fig.~\ref{fig:fig4}(a),(b) in the units of ${\rm eV}_b/A$, where $A$ is the area of a single hexagon of the honeycomb lattice.  We assume that back gate and top gate can change the carrier density globally or locally, respectively, thereby making it possible to tune the Fermi energy $E_F$ of the whole device while  concurrently creating a potential barrier within its smaller region, as demonstrated experimentally~\cite{Young2009}. Without a potential barrier, $U_{\rm tg} = 0$, and in the ballistic regime with no impurities, the only nonzero component in Fig.~\ref{fig:fig4}(a) is  \mbox{$\mathbf{T}^{\rm o} \neq 0$}. Upon introducing \mbox{$U_{\rm tg} \neq 0$} in still clean graphene, Fig.~\ref{fig:fig4}(b) reveals {\em emergence} of \mbox{$\mathbf{T}^{\rm e} \neq 0$}. Furthermore, combined tuning of $E_F$ and potential barrier height in Fig.~\ref{fig:fig4}(b) modulates the ratio \mbox{$|\mathbf{T}^{\rm o}|/|\mathbf{T}^{\rm e}|$} by an order of magnitude, which is of great importance for applications~\cite{Yoon2017}. We fix the height of the homogeneous barrier, within the gray rectangle of the active region region in Fig.~\ref{fig:fig4}(d),(e), at \mbox{$U_{\rm tg} = 0.1$ eV}.

To clarify how $\mathbf{T}^{\rm e}$ emerges from zero value in graphene without any scattering  to nonzero value, Figs.~\ref{fig:fig4}(c) and ~\ref{fig:fig4}(d),(e) plot spatial profiles of nonequilibrium spin densities $\langle \hat{\mathbf{s}}_{i}\rangle_{\rm CD}^\mathrm{e,o}$ [Eq.~\eqref{eq:sot}] in the absence or presence of potential barrier, respectively. In Fig.~\ref{fig:fig4}(c), only $\langle \hat{\mathbf{s}}_{i}\rangle_{\rm CD}^\mathrm{o} \neq 0$ is nonzero along the $y$-axis and slightly out of the plane in Fig.~\ref{fig:sz}(a) for current injected along the $x$-axis. This is the well-known inverse spin-galvanic (or Edelstein) effect~\cite{Edelstein1990,Aronov1989,Offidani2017} in which current through 2D electron system with SOC, and thereby generated spin texture with spin-momentum locking, induces \mbox{$\langle \hat{\mathbf{s}}_{i}\rangle_{\rm CD} \neq 0$}. But here this effect is operative in the ballistic transport regime~\cite{Chang2015,Kalitsov2017}, instead of originally considered diffusive regime~\cite{Edelstein1990,Aronov1989}. Upon introducing impurities  [leading to dashed red line in Fig.~\ref{fig:fig4}(c)], as random on-site potential to establish the diffusive transport regime characterized by the shot noise Fano factor $F \simeq 1/3$~\cite{Lewenkopf2008}, $\langle \hat{\mathbf{s}}_{i}\rangle_{\rm CD}$ acquires additional components in the other two  spatial directions. The came occurs in backscattering from a barrier [leading to solid red line in Fig.~\ref{fig:fig4}(e)]. This means that components of $\langle \hat{\mathbf{s}}_{i}\rangle_{\rm CD}$ emerge along the $x$- [Fig.~\ref{fig:fig4}(c),(e)] and the $z$-axis [Fig.~\ref{fig:sz}] to comprise $\langle \hat{\mathbf{s}}_{i}\rangle_{\rm CD}^\mathrm{e}$ vector. Thus, scattering off either the barrier  or impurities  is required to produce $\langle \hat{\mathbf{s}}_{i}\rangle_{\rm CD}^\mathrm{e}$ and  \mbox{$\mathbf{T}^{\rm e} \neq 0$}. 

\begin{figure}
	\includegraphics[scale=1.0]{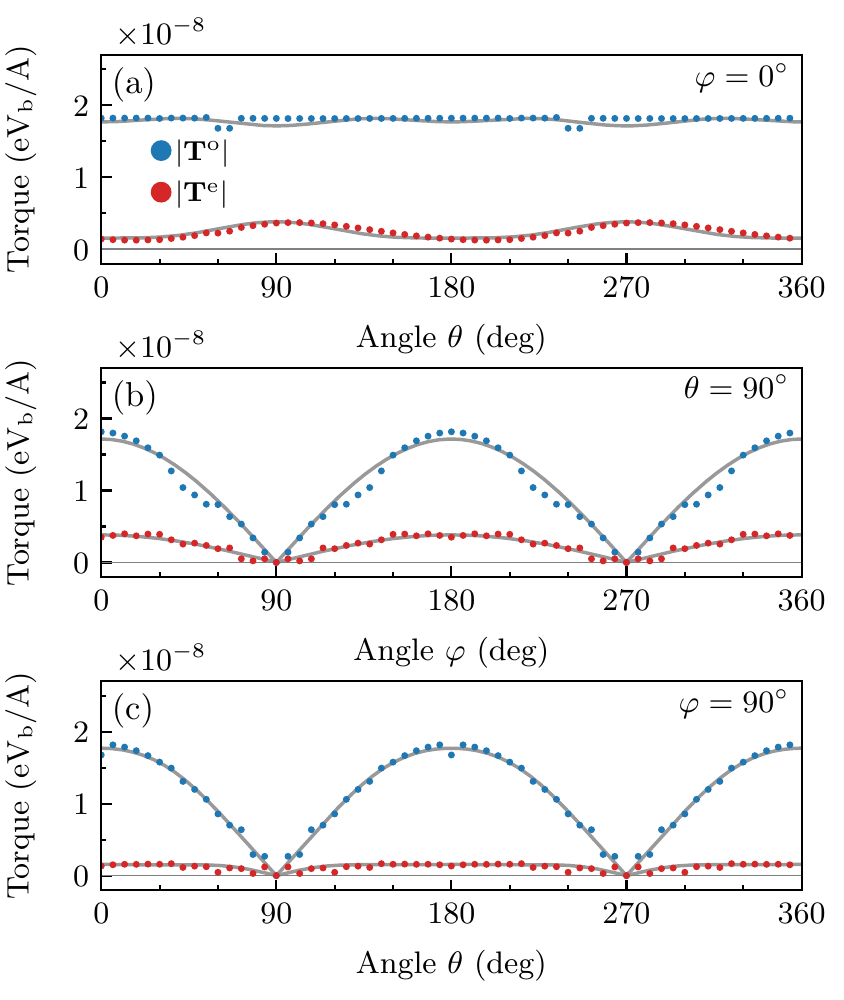}
	\caption{(a)--(c) The angular dependence of SO torque components $|\mathbf{T}^{\rm o}|$ (blue dots) and $|\mathbf{T}^{\rm e}|$ (red dots)  for different orientations of magnetization  $\mathbf{m}_\mathrm{C}$ of proximitized clean graphene at \mbox{$E_F=-0.02$} eV and with potential barrier in the active region due to on-site potential $U_{\rm tg} = 0.1$~eV. Gray lines in the background fit, using Eq.~\eqref{eq:fitsot}, numerically calculated SO torque values (dots). The vector \mbox{$\mathbf{m}_\mathrm{C}=(\sin \theta \cos \varphi, \sin \theta \sin \varphi, \cos \theta)$}  rotates within the $xz$-plane in (a); within the $xy$-plane in (b); and within the $yz$-plane in (c).}
	\label{fig:fig5}
\end{figure}

Previously discussed mechanisms~\cite{Pesin2012a,Qaiumzadeh2015,Ado2017} for purely interfacial \mbox{$\mathbf{T}^{\rm e} \neq 0$} require {\em spin-active} impurities, such as magnetic or SO ones~\cite{Kohno2006}, that scatter spin-$\uparrow$ and spin-$\downarrow$ electrons differently. However, the potential barrier and/or short-ranged impurities we employ in Fig.~\ref{fig:fig4} are {\em spin-independent}. Nonetheless, they can generate {\em skew-scattering}~\cite{Sousa2020,Milletari2017} due to proximity SOC in the band structure  and sublattice-staggered potentials in Eq.~\eqref{eq:hamiltonian} which conspire to tilt [Fig.~\ref{fig:spinmaps_grp_sandwich}] the equilibrium spin textures  out of the plane. This occurs even when $\mathbf{m}_\mathrm{C}=0$, while the barrier also requires $\mathbf{m}_\mathrm{C} \neq 0$. Both cases lead to generation of \mbox{$\langle \hat{\mathbf{s}}_{i}\rangle_{\rm CD}$} components in {\em all spatial directions}. 

This is further clarified by Figs.~\ref{fig:sz}(a) and ~\ref{fig:sz}(b)---complementing Figs.~\ref{fig:fig4}(c) and \ref{fig:fig4}(d),(e), respectively---which reveal the out-of-plane component of CD nonequilibrium spin density  $\langle \hat{s}^{z}_i\rangle_{\rm CD}$. Thus,   spin-dependent skew-scattering~\cite{Milletari2017} generates large out-of-plane component $\langle \hat{s}^z_i \rangle_\mathrm{CD}$  (dashed lines) in Fig.~\ref{fig:sz} which contributes to  $\langle \hat{\mathbf{s}}_i \rangle_\mathrm{CD}^\mathrm{e}$  and the corresponding even component of SO torque, $\mathbf{T}^\mathrm{e}$, via Eq.~\eqref{eq:sot}.


%
\begin{table*}[th]
	\footnotesize 
	\caption{\label{tab:fitT} The fitting parameters in Eq.~\eqref{eq:fitsot}.}
	\begin{ruledtabular}
		\begin{tabular}{c c c c c c c}
			$U_{\rm tg}$ &
			$\tau_0^\mathrm{o}$ &
			$\tau_2^\mathrm{o}$ &
			$\tau_4^\mathrm{o}$ &
			$\tau_0^\mathrm{e}$ &
			$\tau_2^\mathrm{e}$ &
			$\tau_4^\mathrm{e}$ \\
			\footnotesize (eV) &
			\footnotesize (e$\rm V_b$/A) &
			\footnotesize (e$\rm V_b$/A) &
			\footnotesize (e$\rm V_b$/A) &
			\footnotesize (e$\rm V_b$/A) &
			\footnotesize (e$\rm V_b$/A) &
			\footnotesize (e$\rm V_b$/A) \\
			\hline
			\footnotesize $0.0$ &
			\footnotesize $1.112 \times 10^{-8}$ &
			\footnotesize $0.0$ &
			\footnotesize $0.0$ &
			\footnotesize $0.0$ &
			\footnotesize $0.0$ &
			\footnotesize $0.0$ \\
			
			\footnotesize $0.100$ &
			\footnotesize $1.770 \times 10^{-8}$ &
			\footnotesize $2.416 \times 10^{-9}$ &
			\footnotesize $-2.967 \times 10^{-9}$ &
			\footnotesize $1.560 \times 10^{-9}$ &
			\footnotesize $1.234 \times 10^{-11}$ &
			\footnotesize $2.262 \times 10^{-9}$ \\
		\end{tabular}
	\end{ruledtabular}
\end{table*}

To connect $\mathbf{T}^{\rm e}$ and $\mathbf{T}^{\rm o}$ to traditionally discussed DL and FL torque components, we need to take into account typically complex angular dependence of SO torque observed experimentally~\cite{Garello2013} or in first-principles  quantum transport  studies~\cite{Dolui2020,Belashchenko2019,Belashchenko2020}. For this purpose, we fit computational data (dots in Fig.~\ref{fig:fig5}) with an infinite  series~\cite{Dolui2020,Garello2013,Belashchenko2019} for $\mathbf{T}^{\rm e}$ and  $\mathbf{T}^{\rm o}$ vector fields on the unit sphere of orientations of $\mathbf{m}_\mathrm{C}$. The non-negligible terms, compatible with the lattice symmetry, are given by
\begin{subequations}
	\label{eq:fitsot}
	\begin{align}
	\mathbf{T}^{\rm o} = &
	\left[\tau_0^\mathrm{o} +
	\tau_2^\mathrm{o}|\hat{\mathbf{z}} \times\mathbf{m}_\mathrm{C}|^2 + 
	\tau_4^\mathrm{o}|\hat{\mathbf{z}}\times\mathbf{m}_\mathrm{C}|^4
	\right] (\hat{\mathbf{y}} \times \mathbf{m}_\mathrm{C}), \\
	\mathbf{T}^{\rm e}  = &
	\left[\tau_0^\mathrm{e} +
	\tau_2^\mathrm{e}|\hat{\mathbf{z}} \times\mathbf{m}_\mathrm{C}|^2 
	+	
	\tau_4^\mathrm{e}|\hat{\mathbf{z}} \times\mathbf{m}_\mathrm{C}|^4
	\right] \mathbf{m}_\mathrm{C} \times (\hat{\mathbf{y}} \times \mathbf{m}_\mathrm{C}),
	\end{align}
\end{subequations}
The computed angular dependence of even, $\mathbf{T}^{\rm e}$, and odd, $\mathbf{T}^{\rm o}$, SO torque components---as the magnetization vector \mbox{$\mathbf{m}_\mathrm{C}=(\sin \theta \cos \varphi, \sin \theta \sin \varphi, \cos \theta)$} rotates within the $xz$-, $xy$- and $yz$-planes in the setup of Fig.~\ref{fig:fig1} with nonzero potential barrier---is shown by dots in Fig.~\ref{fig:fig5}. Note that for $\theta=0^\circ$, the results in Fig.~\ref{fig:fig5} correspond to the results in Fig.~\ref{fig:fig4}(b)  at \mbox{$E_F = -0.02$ eV}. The functions in Eq.~\eqref{eq:fitsot}, with fitting parameters listed in Table~\ref{tab:fitT}, are then plotted as solid lines in Fig.~\ref{fig:fig5}. Here $\hat{\mathbf{y}}$ and $\hat{\mathbf{z}}$ are the unit vectors along the $y$- and $z$-axis, respectively, for the coordinate system in Fig.~\ref{fig:fig1}.  Note that prior to  fitting, data sets in all three planes ($xz$, $xy$, and $yz$) are joined in a single continuous line for each case (with and without the on-site potential $U_{\rm tg}$), so that they could all be fitted with a single set of fitting parameters. 

In Eq.~\eqref{eq:fitsot} we recognize the lowest-order term \mbox{$\tau_0^\mathrm{o} \hat{\mathbf{y}} \times \mathbf{m}_\mathrm{C}$} as the FL SO torque, which is the {\em only} term (Table~\ref{tab:fitT})  generated in the ballistic transport regime of Fig.~\ref{fig:fig4}(a),(c). The usual DL SO torque~\cite{Manchon2019} is \mbox{$\tau_0^\mathrm{e}\mathbf{m}_\mathrm{C} \times (\hat{\mathbf{y}} \times \mathbf{m}_\mathrm{C})$}. Higher order terms in Eq.~\eqref{eq:fitsot} can have properties of both FL and DL torques~\cite{Belashchenko2020}. Fitted functions  in Eq.~\eqref{eq:fitsot} are also required for using~\cite{Dolui2020} computed SO torque in micromagnetic simulations~\cite{Yan2015a}  of the  dynamics of $\mathbf{m}_\mathrm{C}(t)$. 

\section{Conclusions}\label{sec:conclusions}
In conclusion, using first-principles calculations we design vdW heterostructure where graphene is sandwiched between semiconducting monolayers of ferromagnet Cr$_2$Ge$_2$Te$_6$ and transition-metal  dichalcogenide  WS$_2$ to acquire both SO, of valley-Zeeman and Rashba types, and exchange couplings via the respective proximity effects. Such doubly proximitized graphene offers a playground to investigate different microscopic mechanisms of SO torque. We provide first-principles continuous and tight-binding effective Hamiltonians of doubly proximitized graphene which can be used as a starting point of analytical diagrammatic or computational quantum transport calculations, respectively. In particular, using computational quantum transport, we demonstrate how DL component of SO torque  can be generated {\em solely}  by skew-scattering off {\em  spin-independent} potential barrier or impurities in {\em purely}  two-dimensional electronic transport due to the presence of proximity SOC and its spin texture tilted out-of-plane. This occurs without any spin Hall effect~\cite{Zhu2019a} or three-dimensional transport across the interface~\cite{Kim2017,Amin2018} based mechanisms which are naturally absent from vdW heterostructure.

\acknowledgments
K.~Z. and J.~F. were supported by   the Deutsche Forschungsgemeinschaft (DFG, German Research Foundation) SFB 1277 (Project-ID 314695032) and SPP 1666. 
M.~D.~P. and P.~P. were supported by ARO MURI Award  No.~W911NF-14-0247. K.~D. and B.~K.~N. were supported by DOE Grant  No.~DE-SC0016380. The supercomputing time was provided by  XSEDE, which is supported by NSF Grant No.~ACI-1053575.

\appendix

\section{Computational details of DFT calculations}\label{sec:detailsdft}

The electronic structure calculations and structural relaxation for graphene on Cr$_2$Ge$_2$Te$_6$ , as well as for graphene
sandwiched between WS$_2$ and Cr$_2$X$_2$Te$_6$ with X = \{Si, Ge, Te\}, are performed by DFT  
implemented in {\tt Quantum ESPRESSO} package~\cite{Giannozzi2009}. The heterostructure supercell consists of $5 \times 5$ supercell of graphene 
whose bottom surface is covered by a $\sqrt{3} \times \sqrt{3}$ supercell of Cr$_2$Ge$_2$Te$_6$, while its top surface
is covered by a $4 \times 4$ supercell of WS$_2$.  We stretch the lattice constant of graphene by roughly  2\%---from 2.46~\AA{}~to
$2.5$~\AA{}---and stretch the lattice constant of Cr$_2$Ge$_2$Te$_6$ by roughly 6\%---from $6.8275$~\AA{}~\cite{Carteaux1995} to $7.2169$~\AA{}. The WS$_2$
lattice constant is compressed by roughly 1\%---from 3.153~\AA{}~\cite{Schutte1987} to 3.125~\AA{}. The heterostructure supercell has a lattice constant of $12.5$~\AA{}~and it contains 128 atoms. The distance between WS$_2$ and graphene is $\simeq  3.28$~\AA{}, in agreement with
previous calculations~\cite{Gmitra2016}. The distance between Cr$_2$Ge$_2$Te$_6$ and graphene is about $3.52$~\AA{}, also in
agreement with previous calculations~\cite{Zhang2015a}. 

Bulk vdW crystal Cr$_2$Ge$_2$Te$_6$, which is composed of weakly bound monolayers, has Curie temperature \mbox{$T_\mathrm{C} \simeq 60$ K} and PMA. Each monolayer is formed by edge-sharing CrTe$_6$ octahedra where Ge pairs are located in the hollow sites formed by the octahedra honeycomb. The layers are ABC-stacked, resulting in a rhombohedral $R\overline{3}$ symmetry. 

\begin{figure*}
	\includegraphics[scale=1.0]{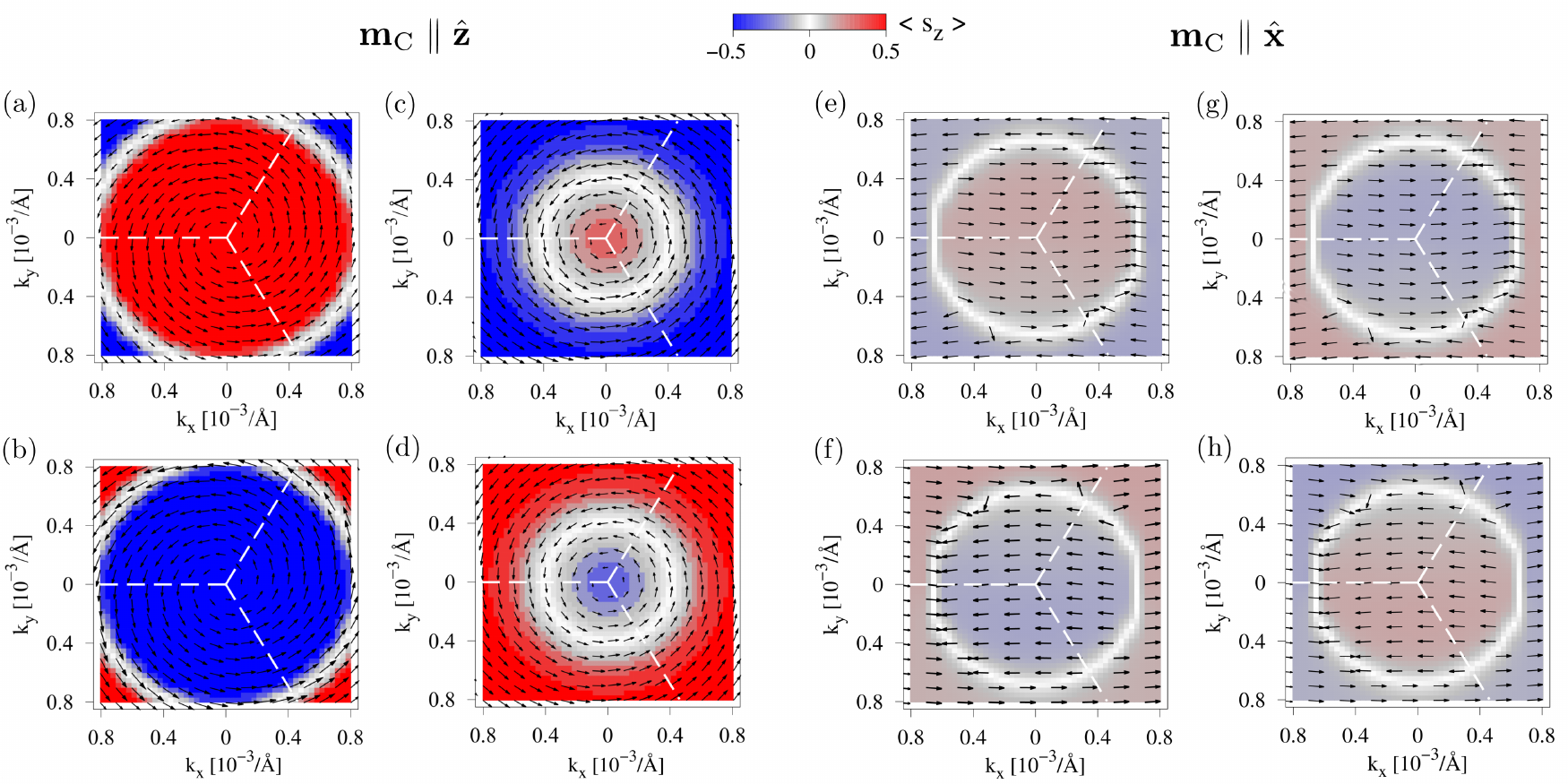}  
	\caption{The DFT-calculated equilibrium spin texture of: (a) conduction; and
		(b) valence band of Cr$_2$Ge$_2$Te$_6$/graphene/WS$_2$ trilayer 
		around the K point. Color corresponds to $\langle \hat{s}_z
		\rangle$ expectation value, while arrows represent $\langle
		\hat{s}_x \rangle$ and $\langle \hat{s}_y \rangle$ expectation values. The
		dashed white lines mark the edges of the Brillouin zone.
		Panels (c) and (d) plot the same information as panels (a) and
		(b), respectively, but around the K$^\prime$ point. Panels (e)--(h) plot the same information as panels (a)--(d), where the proximity 
		magnetization  of graphene $\mathbf{m}_\mathrm{C} \parallel \hat{\mathbf{z}}$ is orthogonal to the plane in (a)--(d) and parallel to the 
		plane  $\mathbf{m}_\mathrm{C} \parallel \hat{\mathbf{x}}$ (i.e., along the $x$-axis in Fig.~\ref{fig:fig1}) in (e)--(h).  
		\label{fig:spinmaps_grp_sandwich}}
\end{figure*}

We use $24\times 24\times 1$ ($18\times 18\times 1$ for the sandwich structure) $k$-point grid in self-consistent calculations to ensure converged results for the proximity exchange interaction and proximity SOC. We also perform open shell calculations that provide the spin-polarized ground state of Cr$_2$Ge$_2$Te$_6$. The Hubbard $U = 1$~eV is used for Cr $d$-orbitals, being in the range of proposed $U$ values for this compound~\cite{Menichetti2019}. The Perdew-Burke-Ernzerhof parametrization~\cite{Perdew1996} of the generalized gradient approximation for the exchange-correlation functional is employed. We use an energy cutoff for charge density of $500$~Ry, and the kinetic energy cutoff for
wavefunctions is $60$~Ry for the scalar relativistic pseudopotential within the projector augmented wave method~\cite{Kresse1999} describing electron-core interactions. When SOC is included, we use the relativistic versions of the pseudopotentials.

For the relaxation of the heterostructures, we add vdW corrections~\cite{Grimme2006, Barone2009} and use quasi-Newton algorithm based on trust radius procedure. In order to simulate quasi-two-dimensional systems, we add a vacuum of \mbox{$20$~\AA{}} which avoids interactions between
periodic images in the slab geometry. To determine the interlayer distances, the atoms of graphene and WS$_2$ are allowed to relax
only along the $z$-axis (perpendicular to the layers), while the atoms of Cr$_2$Ge$_2$Te$_6$ are allowed to move in all directions, until all components of all forces are reduced below $10^{-3}$~Ry/$a_0$ ($a_0$ is the Bohr radius).

\begin{figure}
	\includegraphics[width=0.99\columnwidth]{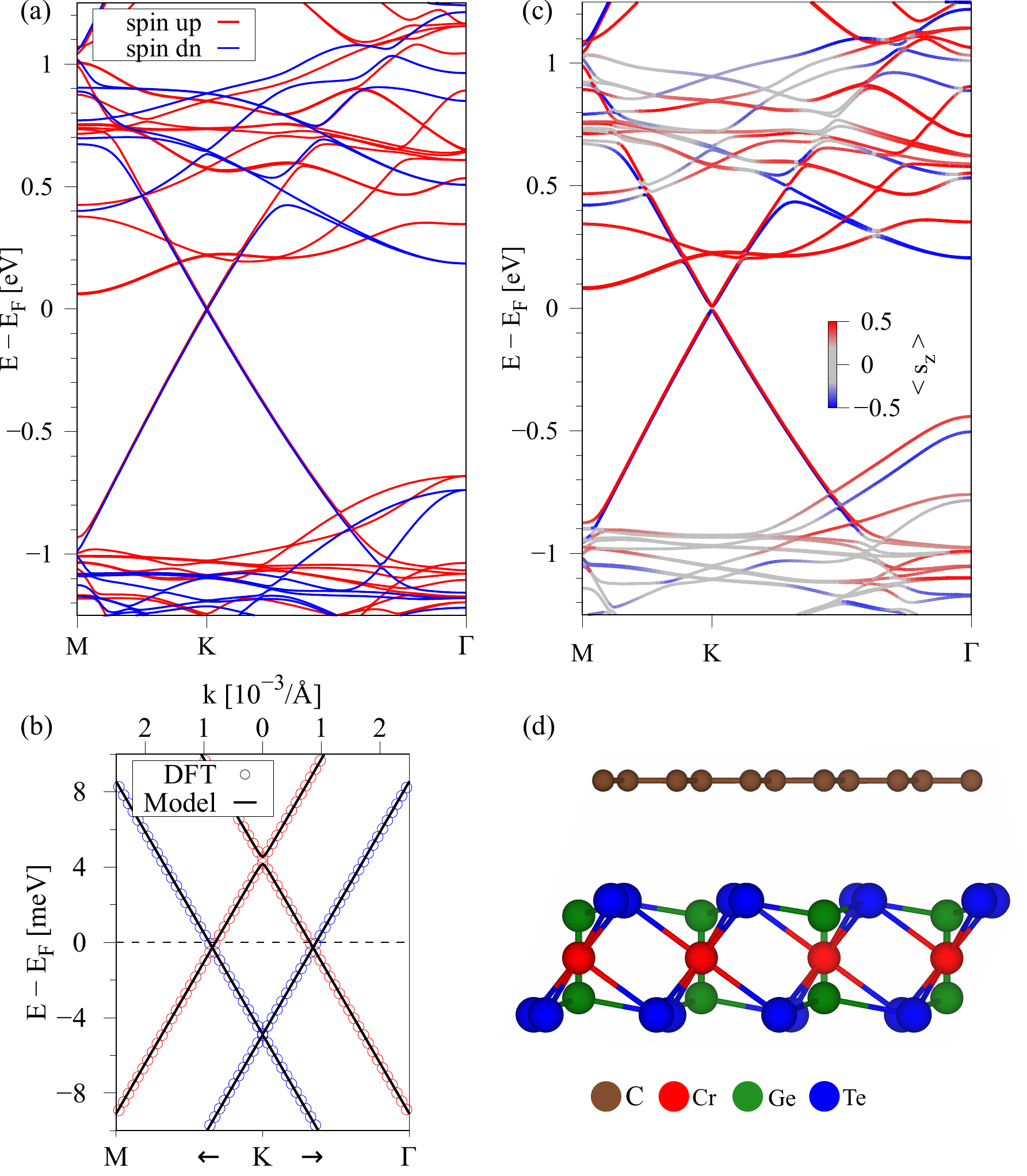}
	\caption{%
		(a) The DFT-calculated electronic band structure of
		Cr$_2$Ge$_2$Te$_6$/graphene bilayer {\em without} SOC. Bands
		with spin $\uparrow$ ($\downarrow$) are shown in red (blue). (b) A zoom to
		DFT-calculated low energy bands (symbols) in panel (a) at the K
		point with a fit to the continuous Hamiltonian (solid lines) in
		Eq.~\eqref{eq:hamiltonian}. (c) DFT-calculated electronic band
		structure of Cr$_2$Ge$_2$Te$_6$/graphene bilayer {\em with} SOC. The color corresponds to $\langle \hat{s}_z \rangle$
		expectation value. (d) Side view of the geometry of monolayer
		graphene on monolayer Cr$_2$Ge$_2$Te$_6$.
		\label{Fig:bands_grp_CGT}}
\end{figure}

\begin{figure}
	\includegraphics[width=0.99\columnwidth]
	{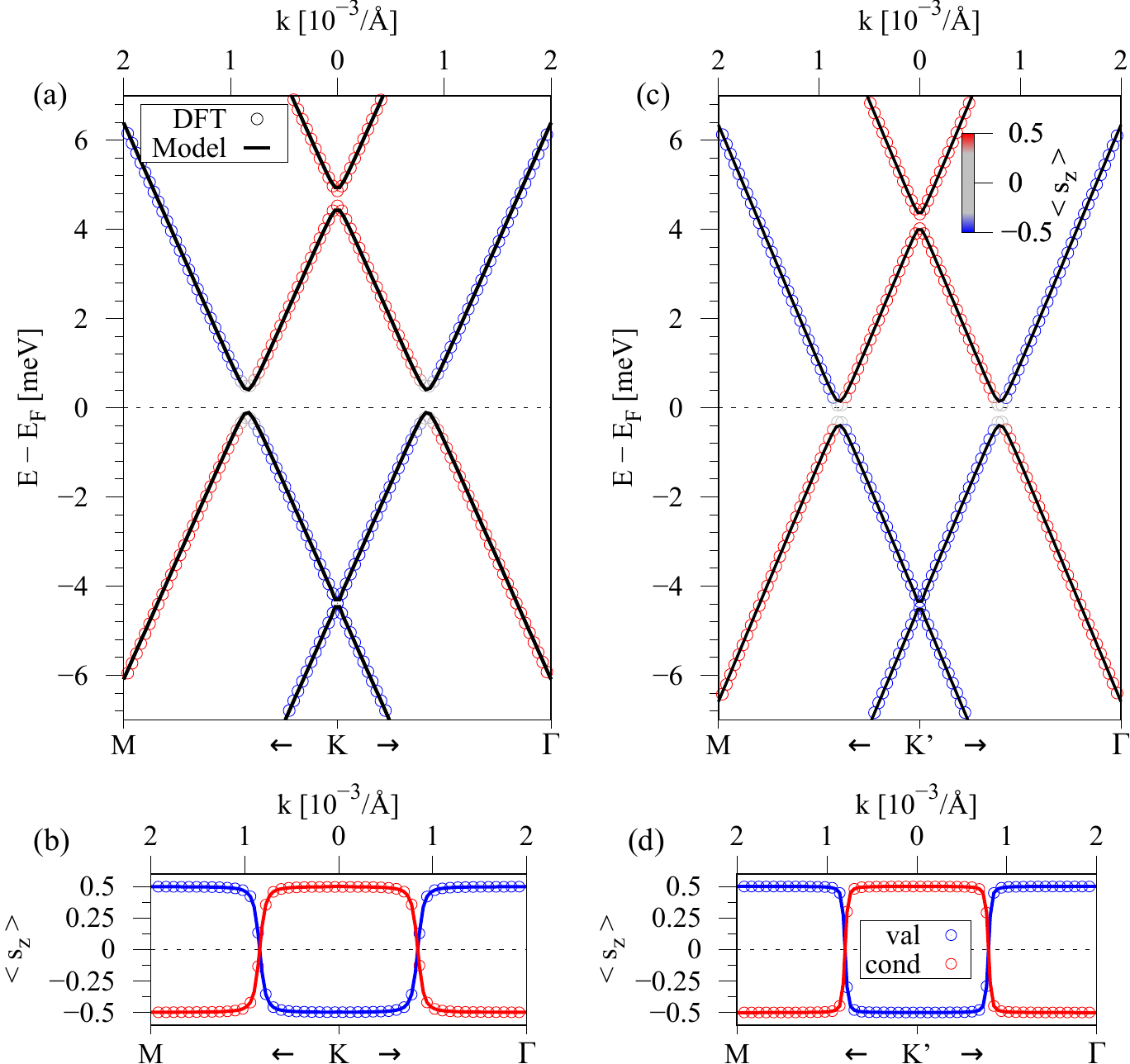}
	\caption{%
		(a) Zoom to DFT-calculated low energy bands in
		Fig.~\ref{Fig:bands_grp_CGT}(c) for Cr$_2$Ge$_2$Te$_6$/graphene
		bilayer around the K point with SOC. Color of the bands
		corresponds to $\langle \hat{s}_z \rangle$ expectation value. (b)
		The $\langle \hat{s}_z \rangle$ expectation value of the valence
		(conduction) band is shown in blue (red). Symbols are DFT-bands and
		solid lines are continuous Hamiltonian [Eq.~\eqref{eq:hamiltonian}] bands. Panels (c) and (d)
		plot the same information as panels (a) and (b), respectively,
		but around the K$^\prime$ point.}
	\label{Fig:bands_grp_CGT_KandKp}
\end{figure}

Table~\ref{tab:table1} summarizes fitting parameters in the continuous Hamiltonian in Eq.~\eqref{eq:hamiltonian} for graphene within Cr$_2$X$_2$Te$_6$/graphene/WS$_2$ trilayers with X = \{Si, Ge, Sn\}. The same parameters are also used in the TB version [Eq.~\eqref{eq:fptbh}] of the continuous Hamiltonian.

\begin{table*}
	\footnotesize
	\caption{\label{tab:table1}
		The fitting parameters of continuous Hamiltonian in
		Eq.~\eqref{eq:hamiltonian} in the  for Cr$_2$X$_2$Te$_6$/graphene/WS$_2$ trilayers   
		with X = \{Si, Ge, Te\}. Here $v_{\textrm{F}}$ is the Fermi
		velocity; $\Delta$ is the staggered potential gap;
		$\lambda_{\textrm{R}}$ is the Rashba SOC strength;
		$\lambda_{\textrm{I}}^\textrm{A}$ and
		$\lambda_{\textrm{I}}^\textrm{B}$ are sublattice-resolved
		intrinsic SOC parameters; $\lambda_{\textrm{ex}}^\textrm{A}$
		and $\lambda_{\textrm{ex}}^\textrm{B}$ are sublattice-resolved
		exchange interaction parameters; $\xi$ is the valley exchange parameter; and $E_\mathrm{D}$ is the Dirac point
		energy. The values of parameters fit the DFT-computed bands
		both with and without SOC.}
	\begin{ruledtabular}
		\begin{tabular}{l c c c c c c c c c c}
			& & $v_{\textrm{F}}$ &
			$\Delta$ &
			$\lambda_{\textrm{ex}}^\textrm{A}$ &
			$\lambda_{\textrm{ex}}^\textrm{B}$ &
			$\lambda_{\textrm{R}}$ &
			$\lambda_{\textrm{I}}^\textrm{A}$ &
			$\lambda_{\textrm{I}}^\textrm{B}$ &
			$\xi$ &
			$E_\mathrm{D}$\\
			calc. & X & $[10^5\frac{\textrm{m}}{\textrm{s}}]$
			& [meV] & [meV] & [meV] & [meV] &
			[meV] & [meV] & [meV] & [meV]\\
			\hline
			no SOC & Si &
			7.921 & 0.865 & -1.899 & -1.811 &
			- & - & - & - & -0.012 \\
			no SOC & Ge &
			7.901 & 1.326 & -3.644 & -3.534 &
			- & - & - & - & -0.054 \\
			no SOC & Sn &
			7.755 & 1.928 & -6.483 & -6.496 &
			- & - & - & - & -0.053 \\
			SOC & Si &
			7.921 & 0.809 & -1.964 & -1.875 &
			-0.397 & 1.127 & -1.152 & 0.268 & -0.758 \\
			SOC & Ge &
			7.903 & 1.252 & -3.481 & -3.650 &
			-0.489 & 1.083 & -1.118 & 0.244 & -0.247\\
			SOC & Sn &
			7.746 & 2.055 & -6.281 & -6.310 &
			-0.696 & 1.009 & -1.057 & 0.022 & -0.610\\
			no SOC \footnotemark[1] & Ge &
			7.979 & 1.602 & -4.591 & -4.422 &
			- & - & - & - & 0.008 \\
			SOC \footnotemark[1] & Ge &
			8.026 & 1.417 & -4.566 & -4.559 &
			-0.467 & 1.148 & -1.184 & 0.158 & 0.004 \\
		\end{tabular}
	\end{ruledtabular}
	\footnotetext[1]{Calculated with {\tt WIEN2k}
		package~\cite{Blaha2001}, using the relaxed geometry from
		{\tt Quantum ESPRESSO} package~\cite{Giannozzi2009},
		and a $k$-point sampling of $12\times 12\times 1$.
		The cutoff is RK$_{\textrm{max}}= 4.0$ and
		the muffin-tin radii are
		R$_{\textrm{Te}} = 2.5$,
		R$_{\textrm{Ge}} = 2.25$, R$_{\textrm{Cr}} = 2.5$,
		R$_{\textrm{C}} = 1.36$, R$_{\textrm{W}} = 2.48$, and
		R$_{\textrm{S}} = 2.03$. The vdW corrections and a Hubbard
		$U = 1.0$ eV are also included.}
\end{table*}

\begin{table*}
	\footnotesize
	\caption{\label{tab:fitresult}
		The fitting parameters of continuous Hamiltonian  in Eq.~\eqref{eq:hamiltonian} for Cr$_2$X$_2$Te$_6$/graphene bilayers with X = \{Si, Ge, Te\}. These parameters have the same meaning as those in Table~\ref{tab:table1}. The values of parameters fit the DFT-computed bands both with and without SOC.}
	\begin{ruledtabular}
		\begin{tabular}{l c c c c c c c c c c}
			& & $v_{\textrm{F}}$ & $\Delta$&
			$\lambda_{\textrm{ex}}^\textrm{A}$ &
			$\lambda_{\textrm{ex}}^\textrm{B}$ &
			$\lambda_{\textrm{R}}$ &
			$\lambda_{\textrm{I}}^\textrm{A}$ &
			$\lambda_{\textrm{I}}^\textrm{B}$ &
			$\xi$ &
			$E_\mathrm{D}$\\
			calc. & X & $[10^5\frac{\textrm{m}}{\textrm{s}}]$
			& [meV] & [meV] & [meV] & [meV] &
			[meV] & [meV] & [meV] & [meV]\\
			\hline
			no SOC & Si &
			8.214 & 0.005 & -1.178 & -1.243 &
			- & - & - & - & 29.424 \\
			no SOC & Ge &
			8.175 & 0.115 & -4.705 & -4.543 &
			- & - & - & - & -0.269 \\
			no SOC & Sn &
			8.066 & 0.119 & -6.995 & -6.894 &
			- & - & - & - & 0.398 \\
			SOC & Si &
			8.219 & 0.062 & -1.195 & -1.166 &
			-0.185 & 0.122 & -0.092 & 0.204 & 10.595 \\
			SOC & Ge &
			8.176 & 0.109 & -4.493 & -4.318 &
			-0.254 & 0.133 & -0.093 & 0.137 & 0.014\\
			SOC & Sn &
			8.069 & 0.170 & -6.830 & -6.722 &
			-0.480 & -0.105 & 0.118 & -0.124 & 0.760\\
			no SOC \footnotemark[1] & Ge &
			8.204 & 0.118 & -5.986 & -5.850 &
			- & - & - & - & 0.001 \\
			SOC \footnotemark[1] & Ge &
			8.213 & 0.114 & -5.338 & -5.216 &
			-0.318 & 0.149 & -0.121 & 0.037 & 0.005 \\
		\end{tabular}
	\end{ruledtabular}
	\footnotetext[1]{Calculated with {\tt WIEN2k}
		package~\cite{Blaha2001}, using a $k$-point sampling of
		$12\times 12\times 1$. The cutoff is RK$_{\textrm{max}}= 4.3$
		and the muffin-tin radii are R$_{\textrm{Te}} = 2.5$,
		R$_{\textrm{Ge}} = 2.25$, R$_{\textrm{Cr}} = 2.5$,
		and R$_{\textrm{C}} = 1.33$. The vdW corrections and a
		Hubbard $U = 1$~eV are also included.}
\end{table*}

\section{Equilibrium spin textures in Cr$_2$Ge$_2$Te$_6$/graphene/WS$_2$ trilayer}\label{sec:spintexture}

Figure~\ref{fig:spinmaps_grp_sandwich} shows DFT-calculated equilibrium spin textures (i.e., expectation values of the Pauli
matrices in the eigenstates of the DFT Hamiltonian) for Cr$_2$Ge$_2$Te$_6$/graphene/WS$_2$ trilayer, with proximity magnetization of graphene pointing out of the plane [Fig.~\ref{fig:spinmaps_grp_sandwich}(a)--(d)] or in the plane [Fig.~\ref{fig:spinmaps_grp_sandwich}(e)--(h)]. The textures are plotted 
in the $k_xk_y$-plane in the vicinity of the K and K$^\prime$ points at energies belonging to the conduction and
valence bands, respectively. Thus, they complement the low-energy band structure shown in Fig.~\ref{fig:fig3}. The bands are
strongly $s_z$-polarized with some small in-plane contribution. The pattern of the in-plane contribution and the corresponding
spin-momentum locking is characteristic of the Rashba SOC-induced spin texture~\cite{Manchon2015}.


\section{Band structure of Cr$_2$Ge$_2$Te$_6$/graphene bilayers}\label{sec:cgtgraphene}

The supercell of Cr$_2$Ge$_2$Te$_6$/graphene bilayer is depicted in
Fig.~\ref{Fig:bands_grp_CGT}(d), where a $5 \times 5$ supercell
of graphene is placed on a $\sqrt{3} \times \sqrt{3}$ supercell
of Cr$_2$Ge$_2$Te$_6$. We keep the lattice constant of graphene
unchanged at $a = 2.46$~\AA{}~and stretch the lattice constant
of Cr$_2$Ge$_2$Te$_6$ by roughly 4\% from
$6.8275$~\AA{}~\cite{Carteaux1995} to $7.1014$~\AA{}. The
resulting supercell has a lattice constant of $12.3$~\AA{}~and
contains 80 atoms in the unit cell. The average distance between
graphene and Cr$_2$Ge$_2$Te$_6$ is relaxed to $3.516$~\AA{}, consistent with
previous calculations~\cite{Zhang2015a}. In the case of
Cr$_2$Si$_2$Te$_6$/graphene and Cr$_2$Sn$_2$Te$_6$/graphene we
get interlayer distances of $3.568$~\AA{}~and $3.541$~\AA{},
respectively.

It has been shown previously~\cite{Zhang2015a} that proximity induced exchange interaction in graphene due to monolayer Cr$_2$Ge$_2$Te$_6$ is an order of magnitude larger than proximity induced SOC by the same monolayer. Therefore, we first perform calculations with SOC turned off.
Figure~\ref{Fig:bands_grp_CGT}(a) shows DFT-calculated band structure for Cr$_2$Ge$_2$Te$_6$/graphene bilayer without SOC. In general, the band structure resembles that of an isolated ferromagnetic semiconductor Cr$_2$Ge$_2$Te$_6$~\cite{Menichetti2019}, with the graphene Dirac cone located in the gap of monolayer Cr$_2$Ge$_2$Te$_6$. The linear dispersion of graphene is nicely preserved, however the bands are spin-split due to proximity induced exchange
interaction.

Without SOC, the band structure around the two valleys K and K$^\prime$ are the same. When SOC is turned off the global band gap
at $E-E_F=0$ is absent. Figure~\ref{Fig:bands_grp_CGT}(c) shows how the band structure gets modified when SOC is turned on. As demonstrated in Fig.~\ref{Fig:bands_grp_CGT}(b), the bands of the continuous Hamiltonian in Eq.~\eqref{eq:hamiltonian}, with parameters provided in Table~\ref{tab:fitresult}, 
fit  perfectly the DFT-calculated low energy bands in a small energy window around the Fermi level.

In addition, Fig.~\ref{Fig:bands_grp_CGT_KandKp} shows a zoom to
the fine structure around the K and K$^\prime$ points, including
a fit with bands of the continuous Hamiltonian in Eq.~(1) in the main
text, when SOC is turned on. The fit can perfectly reproduce the
DFT-computed band structure around both valleys. The fitting 
parameters are summarized in Table~\ref{tab:fitresult} for the
calculations with and without SOC. We find that a tiny gap opens
due to SOC, which is responsible for mixing the two spin
components. The proximity exchange parameters are on the order of
\mbox{$5$~meV}, and proximity SOC is an order of magnitude
smaller. As shown in Table~\ref{tab:fitresult}, DFT calculations
with and without SOC show no significant change of the proximity
exchange parameters $\lambda_{\textrm{ex}}^\textrm{A}$ and
$\lambda_{\textrm{ex}}^\textrm{B}$, the staggered potential
induced gap $\Delta$ and the Fermi velocity $v_{\textrm{F}}$.

In addition to matching the bands,
Fig.~\ref{Fig:bands_grp_CGT_KandKp}(b),(d) demonstrates that
$\langle \hat{s}_z \rangle$ expectation values of the valence and
conduction band agree perfectly with those from the model
Hamiltonian. The bands are fully $s_z$-polarized due to the
strong proximity exchange coupling induced by monolayer
Cr$_2$Ge$_2$Te$_6$. Where SOC mixes the spin-$\uparrow$ and spin-$\downarrow$ 
bands, a gap is opened and $\langle \hat{s}_z \rangle$ expectation
values change in sign. The $\langle \hat{s}_x \rangle$ and $\langle
\hat{s}_y \rangle$ expectation values (not shown) are very small and
show a signature of Rashba SOC in agreement with the small
$\lambda_{\textrm{R}}$ parameter in Table~\ref{tab:fitresult}.

As a further consistency check, we recalculated the band
structure with {\tt WIEN2k} package~\cite{Blaha2001}. The
corresponding fitting parameters are also given in
Table~\ref{tab:fitresult}. The only difference is that the
proximity exchange parameters $\lambda_{\textrm{ex}}^\textrm{A}$
and $\lambda_{\textrm{ex}}^\textrm{B}$ are increased compared to
the calculation with {\tt Quantum ESPRESSO}
package~\cite{Giannozzi2009}, which we can attribute to the
different basis sets, number of $k$-points, and cutoffs used.

\section{Tight-binding version of continuous graphene Hamiltonian}\label{sec:tbh}

The first-principles TB Hamiltonian employed in quantum transport calculations in Figs.~\ref{fig:fig4}, ~\ref{fig:sz}, and ~\ref{fig:fig5} is discretized 
version of the continuous Hamiltonian in Eq.~\eqref{eq:hamiltonian}. Its terms are given by
\begin{subequations}
	\label{eq:fptbh}
	\begin{eqnarray}
		\hat{H} & = & 
		\hat{H}_{0} + 
		\hat{H}_{\Delta} +
		\hat{H}_{\textrm{I}} + 
		\hat{H}_{\textrm{R}} +
		\hat{H}_{\textrm{ex}}, \\
		\hat{H}_{0} & = &
		-t \sum_{\langle i, j\rangle} \sum_{\sigma}
		\hat{c}^\dagger_{i\sigma} \hat{c}_{j\sigma} +
		U_{\rm tg} \sum_{i} \sum_{\sigma}
		\hat{c}^\dagger_{i\sigma} \hat{c}_{i\sigma}, \\
		\hat{H}_{\Delta} & = &
		\sum_{\rm S = A,B}\sum_{i\,\in\,{\rm S}}\sum_{\sigma}
		\Delta \zeta_{\rm S}
		\hat{c}^\dagger_{i\sigma} \hat{c}_{i\sigma},\\
		\hat{H}_{\rm I} & = &
		\sum_{\rm S = A, B}
		\frac{i\lambda_{\rm I}^{\rm S}}{3\sqrt{3}}
		\sum_{\langle\langle i, j \rangle\rangle \in {\rm S}}
		\sum_{\sigma} \nu_{ij}
		{[\hat{s}_z]}_{\sigma\sigma} \hat{c}^\dagger_{i\sigma}
		\hat{c}_{j\sigma}, \\
		\hat{H}_{\rm R} & = &
		\frac{2i\lambda_{\rm R}}{3}
		\sum_{\langle i,j \rangle} \sum_{\sigma \neq \sigma'}
		{[\hat{\mathbf{s}} \times
			{\boldsymbol d}_{ij}]}_{\sigma \sigma'}
		\hat{c}^\dagger_{i \sigma} \hat{c}_{j \sigma'}, \\
		\hat{H}_{\rm ex} & = &
		\sum_{\rm S = A, B}
		(-\zeta_{\rm S})\lambda_{\rm ex}^{\rm S}
		\sum_{i\in {\rm S}} \sum_{\sigma}
		{[\mathbf{m}_{\rm C} \cdot 
			\hat{\mathbf{s}}]}_{\sigma \sigma}
		\hat{c}^\dagger_{i\sigma} \hat{c}_{i\sigma}.
	\end{eqnarray}
\end{subequations}
Here $\hat{c}^\dagger_{i\sigma}$ ($\hat{c}_{i\sigma}$) creates
(annihilates) electron on site $i$ with spin $\sigma=\uparrow, \downarrow$;
\mbox{$t=2.7$ eV} is the nearest-neighbor hopping; $\zeta_{\rm A}
= 1$ for triangular sublattice A and $\zeta_{\rm B} = -1$ for
triangular sublattice B; sum $\langle i,j \rangle$ goes over all
pairs of nearest-neighbor sites, while sum $\langle \langle i,j
\rangle \rangle$ goes over all pairs of next-nearest-neighbor
sites on sublattices A and B; and $\mathbf{d}_{ij}$ is the vector
connecting nearest neighbor sites $i$ and $j$. The parameter $\nu_{ij}$ is either $+1$
or $-1$ depending on the hopping direction.  Other symbols have the same meaning as in Eq.~\eqref{eq:hamiltonian}, 
and the values of the fitting parameters are the same as in Table~\ref{tab:table1}.

The only difference between the continuous Hamiltonian in Eq.~\eqref{eq:hamiltonian} and the TB Hamiltonian in Eq.~\eqref{eq:fptbh} is neglect of   $\hat{H}_{\xi}$ term in Eq.~\eqref{eq:fptbh} due to small value of  $\xi$ parameter in Table~\ref{tab:table1}. Another difference is in that $\mathbf{m}_\mathrm{C}$ as the unit vector along the proximity induced magnetization of graphene in $\hat{H}_{\rm ex}$ can point in any direction 
to make it possible  to study the angular dependence of the SO torque in Fig.~\ref{fig:fig5}. In contrast, this term is fixed only along the $z$-axis 
in Eq.~\eqref{eq:hamiltonian}.

We note that quantum transport calculations using TB  Hamiltonian in Eq.~\eqref{eq:fptbh} are equivalent to computationally much more expensive calculations using DFT Hamiltonian~\cite{Dolui2020}, as long as one remains in the linear-response transport regime driven by small applied bias voltage. The small bias voltage $V_b \ll E_F$ is indeed utilized in the calculations in Figs.~\ref{fig:fig4}, ~\ref{fig:sz} and ~\ref{fig:fig5}. This is due to the fact that self-consistent re-calculation of particle densities and the Hamiltonian due to current flow is nonlinear effect in the bias voltage. Trivially, one should also keep the Fermi energy $E_F$ in Figs.~\ref{fig:fig4} and ~\ref{fig:fig5} within the energy window around the Dirac point where the continuous Hamiltonians in Eq.~\eqref{eq:hamiltonian} or TB Hamiltonian in Eq.~\eqref{eq:fptbh} with parameters in Table~\ref{tab:table1} are applicable.


\begin{thebibliography}{10}
	
	\bibitem{Manchon2019}
	A.~Manchon, I.~M. Miron, T.~Jungwirth, J.~Sinova, J.~Zelezn\'{y}, A.~Thiaville, K.~Garello, and P.~Gambardella, Current-induced spin-orbit torques in	ferromagnetic and antiferromagnetic systems, Rev. Mod. Phys. {\bf 91}, 035004 (2019).
	
	\bibitem{Locatelli2014}
	N.~Locatelli, V.~Cros, and J.~Grollier, Spin-torque building blocks, Nat. Mater. {\bf 13}, 11 (2014).
	
	\bibitem{Ramaswamy2018}
	R.~Ramaswamy, J.~M. Lee, K.~Cai, and H.~Yang, Recent advances in spin-orbit torques: Moving towards device applications, Appl. Phys. Rev. {\bf 5}, 031107 (2018).
	
	
	\bibitem{Borders2017}
	W.~A. Borders, H.~Akima, S.~Fukami, S.~Moriya, S.~Kurihara, Y.~Horio, S.~Sato, and H.~Ohno, Analogue spin-orbit torque device for
	artificial-neural-network-based associative memory operation, Appl. Phys. Expr. {\bf 10}, 013007 (2017).
	
	\bibitem{Wang2017}
	Y. Wang, D. Zhu, Y. Wu, Y. Yang, J. Yu, R. Ramaswamy, R. Mishra, S. Shi, M. Elyasi, K.-L. Teo, Y. Wu, and H. Yang, Room temperature magnetization switching in topological insulator-ferromagnet heterostructures by spin-orbit torques, Nat. Commun. {\bf 8}, 1364 (2017).
	
	\bibitem{Shi2019}
	S.~Shi {\em et al.}, All-electric magnetization switching and Dzyaloshinskii-Moriya interaction in WTe$_2$/ferromagnet heterostructures, Nat. Nanotech. {\bf 14}, 945 (2019).
	
	\bibitem{Zhu2019a}
	L. Zhu, D. C. Ralph, and R. A. Buhrman, Spin-orbit torques in heavy-metal--ferromagnet bilayers with varying strengths of interfacial spin-orbit coupling, Phys. Rev. Lett. {\bf 122}, 077201 (2019).
	
	\bibitem{Dolui2020}
	K.~Dolui, M.~D. Petrovi\'{c}, K.~Zollner, P.~Plech\'{a}\v{c}, J.~Fabian, and B.~K. Nikoli\'{c}, Proximity spin-orbit torque on a two-dimensional magnet within van der Waals heterostructure: Current-driven antiferromagnet-to-ferromagnet reversible nonequilibrium phase transition in bilayer CrI$_3$, Nano Lett. {\bf 20}, 2288 (2020).
	
	\bibitem{Gibertini2019}
	M.~Gibertini, M.~Koperski, A.~F. Morpurgo, and K.~S. Novoselov, Magnetic 2D materials and heterostructures, Nat. Nanotech. {\bf 14}, 408 (2019).
	
	\bibitem{Cortie2019}
	D.~L. Cortie, G.~L. Causer, K.~C. Rule, H.~Fritzsche, W.~Kreuzpaintner, and F.~Klose, Two-dimensional magnets: Forgotten history and recent progress
	towards spintronic applications, Adv. Funct. Mater. {\bf 30}, 1901414  (2019).
	
	\bibitem{Lv2018}
	W.~Lv, Z.~Jia, B.~Wang, Y.~Lu, X.~Luo, B.~Zhang, Z.~Zeng, and Z.~Liu, Electric-field control of spin–orbit torques in WS$_2$/permalloy bilayers, ACS Appl. Mater. Interfaces {\bf 10},  2843 (2018).
	
	\bibitem{Yan2015a}
	S. Yan and Ya. B. Bazaliy, Phase diagram and optimal switching induced by spin Hall effect in a perpendicular magnetic layer, Phys. Rev. B {\bf 91}, 214424 (2015).

	\bibitem{Yoon2017}
	J.~Yoon, S.-W. Lee, J.~H. Kwon, J.~M. Lee, J.~Son, X.~Qiu, K.-J. Lee, and H.~Yang, Anomalous spin-orbit torque switching due to field-like
	torque-assisted domain wall reflection, Sci. Adv. {\bf 3}, e1603099 (2017).
	
	\bibitem{Sinova2015}
	J.~Sinova, S.~O. Valenzuela, J.~Wunderlich, C.~H. Back, and T.~Jungwirth, Spin Hall effects, Rev. Mod. Phys. {\bf 87}, 1213 (2015).

	\bibitem{Dolui2017}
	K.~Dolui and B.~K. Nikoli\'{c}, Spin-memory loss due to spin-orbit coupling at ferromagnet/heavy-metal interfaces: {\em Ab initio} spin-density matrix approach, Phys. Rev. B {\bf 96}, 220403(R)  (2017).
	
	\bibitem{Belashchenko2016}
	K.~D. Belashchenko, A.~A. Kovalev, and M.~van Schilfgaarde, Theory of spin loss at metallic interfaces, Phys. Rev. Lett. {\bf 117}, 207204 (2016).
	
	\bibitem{Gupta2020}
	K. Gupta, R. J.~H. Wesselink, R. Liu, Z. Yuan, and P. J. Kelly, Disorder dependence of interface spin memory loss, Phys. Rev. Lett. {\bf 124}, 087702 (2020). 
	
	\bibitem{Belashchenko2020}
	K.~D. Belashchenko, A.~A. Kovalev, and M.~van Schilfgaarde, Interfacial contributions to spin-orbit torque and magnetoresistance in ferromagnet/heavy-metal bilayers, Phys. Rev. B {\bf 101}, 020407 (2020).
	
	\bibitem{Mahfouzi2020}
	F.~Mahfouzi, R.~Mishra, P.-H. Chang, H.~Yang, and N.~Kioussis, Microscopic origin of spin-orbit torque in ferromagnetic heterostructures: A first-principles approach, Phys. Rev. B {\bf 101}, 060405 (2020).
	
	\bibitem{Luo2019}
	Z. Luo, Q. Zhang, Y. Xu, Y. Yang, X. Zhang, and Y. Wu, Spin-orbit torque in a single ferromagnetic layer induced by surface spin rotation, Phys. Rev. Appl. {\bf 11}, 064021 (2019).
	
	\bibitem{Kim2017} K.-W. Kim, K.-J. Lee, J. Sinova, H.-W. Lee, and M.D. Stiles, Spin-orbit torques from interfacial spin-orbit coupling for
	various interfaces, Phys. Rev. B {\bf 96}, 104438 (2017). 
	
	\bibitem{Amin2018}
	V. P. Amin, J. Zemen, and M. D. Stiles, Interface-generated spin currents, Phys. Rev. Lett. {\bf 121}, 136805 (2018).
	
	\bibitem{Ghosh2018}
	S.~Ghosh and A.~Manchon, Spin-orbit torque in a three-dimensional topological insulator--ferromagnet heterostructure: Crossover between bulk and surface transport, Phys. Rev. B {\bf 97}, 134402 (2018). 
	
	\bibitem{Pesin2012a}
	D. A. Pesin and A. H. MacDonald, Quantum kinetic theory of  current-induced torques in Rashba ferromagnets, Phys. Rev. B {\bf 86}, 014416 (2012).
	
	\bibitem{Qaiumzadeh2015}
	A. Qaiumzadeh, R. A. Duine, and M. Titov, Spin-orbit torques in two-dimensional Rashba ferromagnets, Phys. Rev. B {\bf 92}, 014402 (2015).
	
	\bibitem{Ado2017}
	I. A. Ado, O. A. Tretiakov, and M. Titov, Microscopic theory of spin-orbit torques in two dimensions, Phys. Rev. B {\bf 95}, 094401 (2017).
	
	\bibitem{Kurebayashi2014}
	H. Kurebayashi {\em et al.}, An antidamping spin-orbit torque originating from the Berry curvature, Nat. Nanotechnol. {\bf 9}, 211 (2014).
	
	\bibitem{Menichetti2019}
	G.~Menichetti, M.~Calandra, and M.~Polini, Electronic structure and magnetic properties of few-layer Cr$_2$Ge$_2$Te$_6$: The key role of nonlocal electron-electron interaction effects, 2D Mater. {\bf  6}, 045042 (2019).
	
	\bibitem{Carteaux1995}
	V.~Carteaux, D.~Brunet, G.~Ouvrard, and G.~Andre, Crystallographic, magnetic and electronic structures of a new layered ferromagnetic compound Cr$_2$Ge$_2$Te$_6$, J. Phys.: Condens. Mat. {\bf 7}, 69 (1995).
	
	\bibitem{Gmitra2016}
	M.~Gmitra, D.~Kochan, P.~H\"ogl, and J.~Fabian, Trivial and inverted Dirac bands and the emergence of quantum spin Hall states in graphene on transition-metal dichalcogenides, Phys. Rev. B {\bf 93}, 155104 (2016).
	
	\bibitem{Gmitra2009}
	M.~Gmitra, S.~Konschuh, C.~Ertler, C.~Ambrosch-Draxl, and J.~Fabian, Band-structure topologies of graphene: Spin-orbit coupling effects from first principles, Phys. Rev. B {\bf 80}, 235431 (2009).

	
	\bibitem{Sousa2020}
	F. J. Sousa, G. Tatara, A. Ferreira, Emergent spin-orbit torques in two-dimensional material/ferromagnet interfaces, {\tt arXiv:2005.09670} (2020).
	
	\bibitem{Freimuth2014}
	F.~Freimuth, S.~Bl\"ugel, and Y.~Mokrousov, Spin-orbit torques in Co/Pt(111) and Mn/W(001) magnetic bilayers from first principles, Phys. Rev. B {\bf 90}, 174423 (2014).
    
	\bibitem{Nikolic2018}
	B.~K. Nikoli\'{c}, K.~Dolui, M.~Petrovi\'{c}, P.~Plech\'{a}\v{c}, T.~Markussen, and K.~Stokbro, First-principles quantum transport modeling of spin-transfer and spin-orbit torques in magnetic multilayers, in  {\em Handbook of Materials Modeling: Applications:  Current and Emerging Materials}, edited by W. Andreoni and  S. Yip (Springer, Cham, 2018); {\tt arXiv:1801.05793}.
	
	\bibitem{Dolui2020a}
	K. Dolui and B. K. Nikoli\'{c}, Spin-orbit-proximitized ferromagnetic metal by monolayer transition metal dichalcogenide: Atlas of spectral functions, spin textures and spin-orbit torques in Co/MoSe$_2$, Co/WSe$_2$ and Co/TaSe$_2$ heterostructures, {\tt arXiv:2006.11335} (2020).
	
	\bibitem{Belashchenko2019}
	K.~D. Belashchenko, A.~A. Kovalev, and M.~van Schilfgaarde, First-principles calculation of spin-orbit torque in a Co/Pt bilayer, Phys. Rev. Mater. {\bf 3}, 011401 (2019).
	
	\bibitem{Stefanucci2013}
	G.~Stefanucci and R.~van Leeuwen, \emph{Nonequilibrium Many-Body Theory of Quantum Systems: A Modern Introduction}
	(Cambridge University Press, Cambridge, 2013).
	
	\bibitem{Chang2015}
	P.-H. Chang, T.~Markussen, S.~Smidstrup, K.~Stokbro, and B.~K. Nikoli\'{c}, Nonequilibrium spin texture within a thin layer below the surface of current-carrying topological insulator ${\mathrm{Bi}}_{2}{\mathrm{Se}}_{3}$: A first-principles quantum transport study, Phys. Rev. B {\bf 92}, 201406(R) (2015).
	
	\bibitem{Kalitsov2017}
	A. Kalitsov, S. A. Nikolaev, J. Velev, M. Chshiev, and O. Mryasov, Intrinsic spin-orbit torque in a single-domain
	nanomagnet, Phys. Rev. B {\bf 96}, 214430 (2017).
	
	\bibitem{Young2009}
	A.~F. Young and P.~Kim,  Quantum interference and Klein tunnelling in graphene heterojunctions, Nat. Phys. {\bf  5}, 222 (2009).
	
	\bibitem{Liu2012d}
	M.-H. Liu and K.~Richter, Efficient quantum transport simulation for bulk graphene heterojunctions, Phys. Rev. B {\bf 86}, 115455 (2012).
	
	\bibitem{Milletari2017}
	M. Milletar\`{\i}, M. Offidani, A. Ferreira, and R. Raimondi, Covariant conservation laws and the spin Hall effect in Dirac-Rashba systems, Phys. Rev. Lett. {\bf 119}, 246801 (2017).	
	
	\bibitem{Giannozzi2009}
	P.~Giannozzi {\em et al.}, 
	QUANTUM ESPRESSO: A modular and open-source software project
	for quantum simulations of materials,
	J. Phys.: Condens. Mat. {\bf 21}, 395502 (2009).
	
	\bibitem{Blaha2001}
	P. Blaha, K. Schwarz, G. K. H. Madsen, D. Kvasnicka, J. Luitz, R. Laskowski, F. Tran, and L. D. Marks, {\em WIEN2k: An Augmented Plane Wave Plus Local Orbitals Program for Calculating Crystal Properties} (Vienna University of Technology, Vienna, 2019); ISBN 3-9501031-1-2
	
	\bibitem{Phong2017}
	V.~T. Phong, N.~R. Walet, and F.~Guinea, Effective interactions in a graphene layer induced by the proximity to a ferromagnet, 2D Mater. {\bf 5}, 014004  (2017).
	
	\bibitem{Kochan2017}
	D.~Kochan, S.~Irmer, and J.~Fabian, Model spin-orbit coupling Hamiltonians for graphene systems, Phys. Rev. B {\bf 95}, 165415 (2017).
	
	\bibitem{Marmolejo-Tejada2017}
	J.~M. Marmolejo-Tejada, P.-H. Chang, P.~Lazi\'{c}, S.~Smidstrup, D.~Stradi, K.~Stokbro, and B.~K. Nikoli\'{c}, Proximity band structure and spin textures on both sides of topological-insulator/ferromagnetic-metal interface and their charge transport probes, Nano Lett.  {\bf 17}, 5626 (2017).
	
	\bibitem{Zutic2019}
	I.~\v{Z}uti\'{c}, A.~Matos-Abiague, B.~Scharf, H.~Dery, and K.~Belashchenko, Proximitized materials, Mater. Today {\bf 22}, 85 (2019). 
	
	\bibitem{Gmitra2015}
	M.~Gmitra and J.~Fabian, Graphene on transition-metal dichalcogenides: A platform for proximity spin-orbit physics and optospintronics,
	Phys. Rev. B {\bf 92}, 155403 (2015).
	
	\bibitem{Hallal2017}
	A.~Hallal, F.~Ibrahim, H.~Yang, S.~Roche, and M.~Chshiev, Tailoring magnetic insulator proximity effects in graphene: First-principles calculations, 2D Mater.  {\bf 4}, 025074 (2017).
	
   
    \bibitem{Zhang2015a}
   J.~Zhang, B.~Zhao, Y.~Yao, and Z.~Yang, Robust quantum anomalous Hall effect in graphene-based van der Waals heterostructures, Phys. Rev. B {\bf 92}, 165418 (2015).
   	
	\bibitem{Garello2013}
	K. Garello, I. M. Miron, C. O. Avci, F. Freimuth, Y. Mokrousov, S. Bl\"{u}gel, S. Auffret, O. Boulle, G. Gaudin, and P. Gambardella, Symmetry and magnitude of spin-orbit torques in ferromagnetic heterostructures, Nat. Nanotech. {\bf 8}, 587 (2013).
	
	\bibitem{Edelstein1990} 
	V. Edelstein, Spin polarization of conduction electrons induced by electric current in two-dimensional asymmetric electron system,
	Solid State Commun. {\bf 73}, 233 (1990).
	
	\bibitem{Aronov1989} A. G. Aronov and Y. B. Lyanda-Geller, Nuclear electric  resonance and orientation of carrier spins by an electric field, JETP Lett. {\bf 50}, 431 (1989).
	
	\bibitem{Offidani2017}
	M. Offidani, M. Milletarì, R. Raimondi, and A. Ferreira, Optimal charge-to-spin conversion in graphene on transition-metal dichalcogenides, Phys. Rev. Lett. {\bf 119}, 196801 (2017).
	
	\bibitem{Lewenkopf2008}
	C. H. Lewenkopf, E. R. Mucciolo, and A. H. Castro Neto, Numerical studies of conductivity and Fano factor in disordered graphene, Phys. Rev. B {\bf 77},  
	081410(R) (2008).
	
    \bibitem{Kohno2006}
    H. Kohno, G. Tatara, and J. Shibata,  Microscopic calculation of spin torques in disordered ferromagnets, J.  Phys. Soc. Japan {\bf 75}, 113706 (2006).
    
    \bibitem{Schutte1987}
    W.~J. Schutte, J.~L. De Boer, and F.~Jellinek, Crystal structures of tungsten disulfide and diselenide, J. Solid State Chem. {\bf 70}, 207 (1987).
    
    
    \bibitem{Perdew1996}
    J.~P. Perdew, K.~Burke, and M.~Ernzerhof, Generalized gradient approximation made simple, Phys. Rev. Lett. {\bf 77}, 3865 (1996).
    
    \bibitem{Kresse1999}
    G.~Kresse and D.~Joubert,
    From ultrasoft pseudopotentials to the projector augmented-wave method, Phys. Rev. B {\bf 59}, 1758 (1999).
    
    \bibitem{Grimme2006}
    S.~Grimme, Semiempirical GGA-type density functional constructed with a long-range dispersion correction, J. Comput. Chem. {\bf 27}, 1787 (2006).
    
    \bibitem{Barone2009}
    V.~Barone, M.~Casarin, D.~Forrer, M.~Pavone, M.~Sambi, and A.~Vittadini, Role and effective treatment of dispersive forces in materials: Polyethylene and graphite crystals as test cases, J. Comput. Chem. {\bf 30}, 934 (2009).
    
    \bibitem{Manchon2015}
    A. Manchon, H. C. Koo, J. Nitta, S. M. Frolov, and R. A. Duine, New perspectives for Rashba spin-orbit coupling, Nat. Mater. {\bf 14}, 871 (2015).
    
\end{thebibliography}
\end{document}